%% file: main.tex
\documentclass[twocolumn]{aastex631}

\usepackage{amsmath,amstext}
\usepackage{gensymb}
\usepackage{subfigure}
\usepackage{xcolor}
\usepackage[T1]{fontenc}

\shorttitle{ACES Compact Source Catalog}
\shortauthors{Wallace et al.}

\begin{document}

\title{ACES VII. Compact Continuum Source Catalog of the Central Molecular Zone}

\input{authors}
\begin{abstract}

The Central Molecular Zone (CMZ) resides in the inner few hundred parsecs of our Galaxy, and despite being the largest reservoir of dense molecular gas in the Milky Way, it has a relatively low present-day star formation rate (SFR) of $\sim0.08~M_{\odot}~\text{yr}^{-1}$. Continuum and spectral line observations from the Atacama Large Millimeter/submillimeter Array (ALMA) CMZ Exploration Survey (ACES) provide the first full-coverage, high-resolution map of the inner 200 parsecs of the CMZ at 3 mm. In this paper we present the ACES catalog of compact continuum sources, the most complete catalog of potential sites of star formation in the CMZ to date. Using an automated dendrogram-based source extraction procedure in combination with a by-eye morphological classification scheme, we produce a `full' catalog of 1735 detections in total. Additionally, we use spectral index measurements to generate a `filtered' catalog of 567 sources with minimal contamination from non-thermal filaments and extended free-free emission. We find that 359 ($\sim63\%$) of the filtered catalog sources are located at column densities $< 10^{23}$ cm$^{-2}$, outside of the densest molecular cloud regions, 195 of which have not been identified in previous surveys. After cross-referencing with various catalogs generated from data at different wavelengths, we consider it likely that many of these newly discovered detections are produced by pre/protostellar sources or compact H\textsc{ii} regions.

\end{abstract}
\keywords{Galactic center (565)}

\section{Introduction} \label{sec:intro}

The Central Molecular Zone (CMZ) is the enormous accumulation of dense molecular gas ($\sim 2^{+2}_{-1} \times 10^{7}$ M$_{\odot}$) contained within the inner few hundred parsecs ($R\simeq 300$ pc) of the Milky Way \citep{morris_serabyn_1996, Henshaw_2023, Battersby_2025a}. The interstellar medium (ISM) of the CMZ is characterized by conditions that are extreme in comparison to that of the Solar neighborhood, with measured gas temperatures \citep{Ginsburg_2016, Immer_2016}, densities \citep{Mills_2018}, and velocity dispersions \citep{Shetty_2012,Kauffman_2017a} that are at least an order of magnitude greater than that of gas in the Galactic disk at similar size scales. Additionally, the clouds in the CMZ have a unique gas chemistry when compared to nearby clouds \citep{Belloche_2016, Belloche_2025, Moller_2021, Colzi_2022} and are subject to magnetic fields with strengths that exceed those in the local ISM \citep{Pillai_2015, Lu_2024, Zhao_2025}. Overall, these conditions generate an environment that is considerably different than those found in local star forming regions, and is instead more similar to high redshift galaxies \citep{Kruijssen_Longmore_2013}. 

In general, the total star formation rate (SFR) in molecular clouds is linearly related to its dense molecular gas mass, a trend known as the ``dense gas star formation relation'' \citep{Gao_Solomon_2004}. Intriguingly, the current SFR estimate for the CMZ is an order of magnitude lower than expected from this relation, which may imply an increased surface density threshold for star formation in the CMZ \citep{Longmore_2013, Barnes_2017}. 

This dearth in star formation may be explained by the extreme environmental conditions present in the CMZ or potentially by temporal stochasticity in the star formation in this region, with episodic starbursts and periods of quiescence. Age estimates for stellar populations in the CMZ \citep{Nogueras-Lara_2020} as well as some hydrodynamical simulations \citep{Krumholz_Kruijssen_2015, Torrey_2017, 2019_Armillotta, Tress_2020, Sormani_2020, Tress_2025} support this theory of an irregular star formation history. In addition, the turbulence in the CMZ may be dominated by solenoidal driving due to strong shearing motions, which can reduce the SFR by about an order of magnitude compared to compressive driving of turbulence, with the latter being more common across the rest of the Galactic disk \citep{Federrath_Klessen_2012, Federrath_2016, Gerrard_Federrath_2026}.

Much of our current understanding of the cold, dense gas and dust capable of forming stars in the CMZ comes from far-infrared (FIR) \citep{Molinari_2011} and millimeter wavelength \citep{Bally_2010, Aguirre_2011, Ginsburg_2013, Pound_2018, Tang_2021} surveys covering the entire CMZ; however, they lack the spatial resolution needed to identify individual embedded sites of star formation. In recent years, there have been several follow-up interferometric observations exploring Galactic center star formation at the scale of dense (pre)star-forming cores ($\sim0.01-0.1$ pc) \citep[e.g.][]{Sanchez_monge_2017, Walker_2018, Ginsburg_2018, Lu_2019a,  Lu_2020}. These studies uncovered compact substructures within some of the most prominent CMZ clouds, and have shown that there is an overall lack of dense, compact sources that are actively forming stars. These findings are consistent with what was found in the Submillimeter Array (SMA) CMZoom survey, which was the first millimeter wavelength (1.3 mm) survey with complete coverage of the dense gas (N(H$_{2}) > 10^{23}$ cm$^{-2}$) contained within the CMZ \citep{Battersby_2020}. These observations revealed an overall lack of compact structures, contributing less than 10\% of the total cloud mass for almost all of the identified clouds in the CMZ, excluding only the most active sites of star formation located in Dust Ridge
cloud C, Sgr B2, and SgrC. The total CMZ SFR estimated from the CMZoom compact continuum source catalog is $\sim 0.08~$M$_{\odot}$/yr \citep{Hatchfield_2020, Hatchfield_2024}, similar to SFR estimates made in previous studies using alternative methods \citep[e.g.][]{Barnes_2017}. 

Although the CMZoom survey has produced the most complete catalog of embedded sites of star formation in the CMZ to date, it does not capture compact sources located outside of the densest molecular clouds. This lack of coverage may result in the underestimation of the incipient SFR in the Galactic center. To obtain a more complete understanding of the global SFR, star formation efficiency (SFE) and the star formation history in this region, it is important to survey the full area of the CMZ at high spatial resolution. The ALMA CMZ Exploration Survey (ACES) \citep{Longmore_2026} is the first millimeter wavelength survey that uniformly observes the CMZ at sub-parsec spatial resolution. With the data from this unprecedented survey, we are able to resolve the scales of star formation towards the entire inner 200 pc of the CMZ for the first time. 

In this paper we present the first compact continuum source catalog covering the a large fraction of the CMZ using 3 mm continuum observations from ACES \citep{Ginsburg_2026}. In Section \ref{sec:data} we summarize the properties of the ACES continuum observations, as well as information on the ancillary data that we use. In Section \ref{sec:methods} we describe our dendrogram-based source extraction procedure, the by-eye classification method used to generate the ACES catalog, and our method for cross-matching with catalogs at other wavelengths. Afterwards, we provide initial insight into the fundamental characteristics of our sources, as well as their spatial distribution, in Section \ref{sec:results}. In Section \ref{sec:discussion} we compare source properties with those found in the CMZoom survey, and discuss trends in the distribution of sources and speculate on their nature. Finally, we summarize our conclusions in Section \ref{sec:conclusions}.

\begin{figure*}
\begin{centering}
\includegraphics[width=\textwidth]{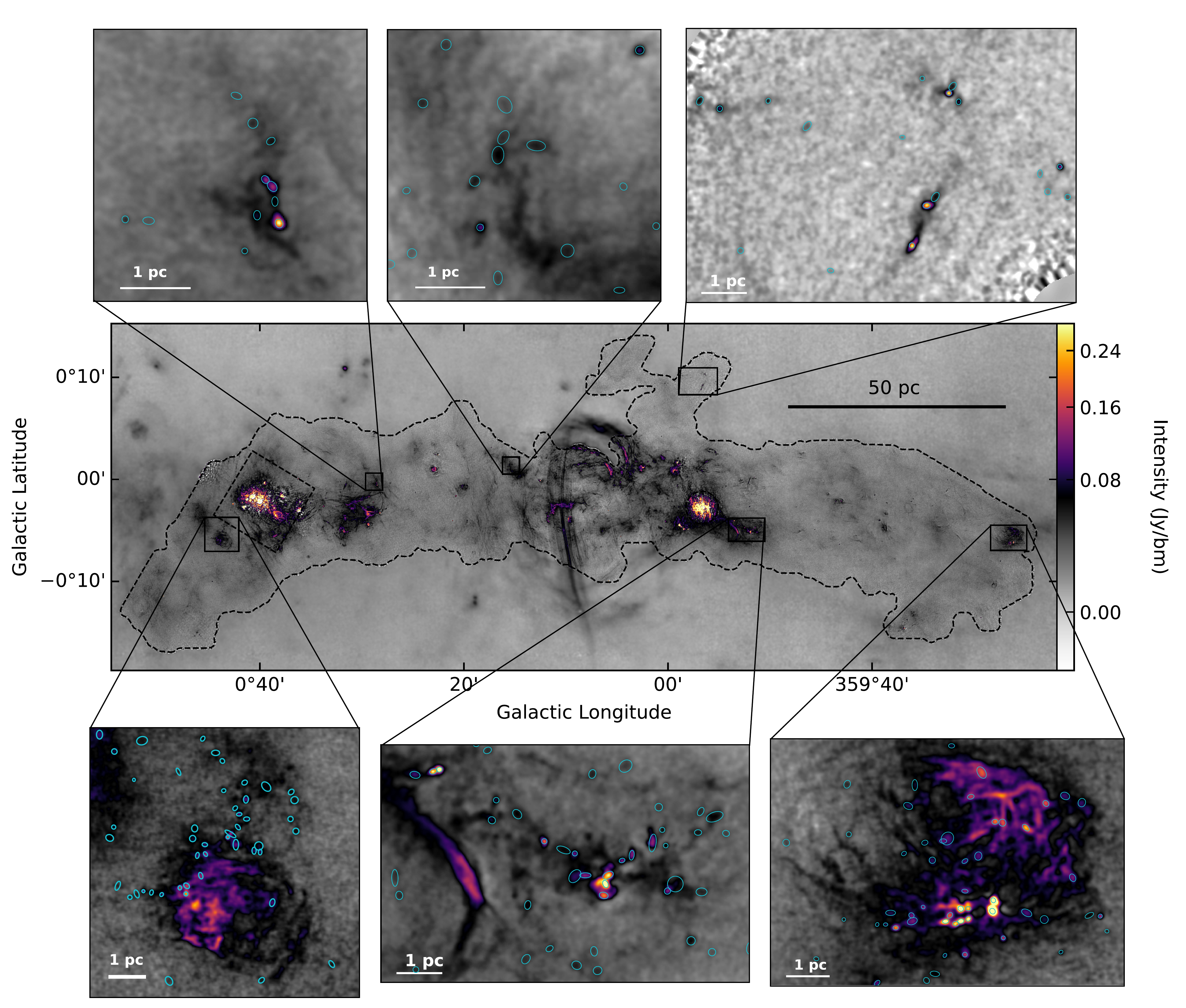}

\caption{An overview of the CMZ as seen at 3 mm with the feathered ACES continuum mosaic. Included are zoom-in panels of specific regions of interest throughout the CMZ. The black, dashed contours show the area of the CMZ that the ACES catalog covers. Sources found within the rectangular and circular regions are excluded from the catalog due to the impact of interferometric imaging artifacts. The cyan ellipses represent the TGIF 2D Gaussian fits for each of the cataloged sources from the ``filtered'' ACES catalog, with FWHM major and minor axes. A white 1 pc scalebar is included in the bottom left corner of each zoom-in panel. The full mosaic image has a square root color stretch for the inner 99.9\% of the pixel values.} 
\label{fig:aces_mosiac}
\end{centering}
\end{figure*}

\section{Data} \label{sec:data}

\subsection{ACES ALMA Observations} \label{subsec:alma_obs}

The ALMA CMZ Exploration Survey (ACES) is an Atacama Large Millimeter/submillimeter Array (ALMA) Band 3 Large Program \citep{Longmore_2026} with coverage of the inner 200 pc of the CMZ. This survey contains 3 mm continuum and spectral line observations that enable the tracing of dusty star forming cores, dense gaseous filaments, and shocked molecular gas. ACES covers a total area of 1286 square arcminutes, with a typical angular resolution of $\sim1.5-2.5$\arcsec, which corresponds to $\sim0.05-0.1$ pc at the assumed Galactic center distance used throughout this work of 8178 pc \citep{grav_collab_2019}. 

Although our analysis uses only the ACES continuum observations presented in \cite{Ginsburg_2026}, detailed information on the ACES spectral line observations can be found in \cite{Walker_2026}, \cite{Lu_2026}, and \cite{hsieh_2026}.

\subsection{ACES 3 mm Continuum Mosaic} \label{subsec:cont_data}

For the results presented in this paper, we use the ACES continuum observations made using the ALMA 12m array, as this configuration is most sensitive to small spatial scales, making it ideal for compact source extraction. The ACES continuum map has a circular FWHM beam size of 2.56$\arcsec$ ($\sim$ 0.1 pc) and is sensitive to size scales up to 10-20$\arcsec$ ($\sim0.4-0.8$ pc). We note that this angular resolution is slightly lower than the resolution of 2.30$\arcsec$  reported in \cite{Ginsburg_2026}. The 2.30$\arcsec$ version of the mosaic was created after we implemented the manual source removal phase of our cataloging procedure (see Section \ref{subsec:manual_removal}). However, this minor update in angular resolution does not impact the conclusions presented in this work.

The 45 fields in the ACES 12m continuum have variations in \textit{uv}-coverage that produces non-uniform noise across the mosaic. To account for this in our source extraction procedure, we use an ACES noise map produced using the method outlined in \cite{Ginsburg_2026}\footnote{The noise map we use in this paper was made using the 2.56\arcsec ~resolution data and not the publicly available ACES noise map, which was generated from the 2.3\arcsec ~resolution data.}. To create this map, a smoothed version of the continuum mosaic is generated by convolving with a Gaussian kernel with a width three times that of the synthesized beam. Then, a residual image is produced by subtracting the smoothed image from the original image. This residual map is then squared to create the variance map, which is then convolved with the same
Gaussian kernel. The final RMS map is produced by taking the square root of the smoothed variance map and then masking out pixels above 2.5$\sigma$ in the original mosaic. This procedure is repeated until no additional pixels are added to the mask. This approach creates a noise map that reduces the contribution of point sources, and is an appropriate estimate for spatial scales similar to the beam. In our source extraction procedure described in Section \ref{subsec:dendro_methods}, we scale our dendrogram, the hierarchical clustering algorithm used to create the catalog, by a global noise value, $\sigma_{\text{global}}$, which we take to be the median value of the noise map, $\sim 100~\mu$Jy/beam. 

We also use the ACES continuum map that has been combined with 3mm single-dish observations from the MUSTANG-2 Galactic Plane Survey (MGPS90) \citep{Ginsburg_2020} and the Three millimeter Expanded Nucleus Survey (TENS) \citep{Ginsburg_2026}. For the purposes of this paper, we only use the single-dish combined mosaic for visualization purposes and for generating a spectral index map as described in Section \ref{subsec:mw_assoc}. We do not measure any compact continuum source properties with it. 

For a more detailed description of the continuum data processing, single-dish combination, and noise map creation, we refer the reader to the ACES Continuum Imaging Paper \citep{Ginsburg_2026}.

\subsection{Ancillary Data}
\label{other_data}

\subsubsection{The 1.28 GHz MeerKAT Galactic Center Mosaic}
\label{meerkat_data}
In our by-eye catalog pruning procedure (see Section \ref{subsec:dendro_methods}), we use 1.28 GHz continuum observations made using the MeerKAT radio telescope \textit{L}-Band receivers (846-1712 MHz) \citep{Heywood_2022} to determine which sources in our catalog are associated with compact objects observed at radio wavelengths. The MeerKAT total-intensity mosaic is comprised of 20 pointings, covering the inner $3.5\degree\times2.5\degree (\textit{l}\times\textit{b})$ of the Milky Way Galactic center and encompasses the ACES footprint completely. The mosaic has a circular synthesized beam size of $\sim 4\arcsec$ ($\sim0.16$ pc at an assumed Galactic center distance of 8178 pc). For a more detailed description of the MeerKAT data processing, we refer the reader to \cite{Heywood_2022}. 

\subsubsection{The CMZoom Survey: 1.3 mm Continuum Mosaic and Compact Continuum Source Catalog}
\label{cmzoom_cat}

The CMZoom Survey \citep{Battersby_2020} is a Submillimeter Array (SMA) 1.3 mm survey with complete coverage of regions in the CMZ with a molecular hydrogen column density $\geq 10^{23}$ cm$^{-2}$. The CMZoom continuum observations have a median rms of 13 mJy beam$^{-1}$ and a typical angular resolution of $\sim3$\arcsec, corresponding to a linear resolution of $\sim0.1$ pc at a distance of 8.178 kpc.

In this paper, we make comparisons between our cataloged sources and those identified in the CMZoom compact continuum source catalog \citep{Hatchfield_2020}. This catalog is composed of compact sources detected in the SMA continuum mosaic. These sources were extracted from the SMA continuum data using a pruned dendrogram algorithm, taking local noise variations into consideration. The high-completeness version of the CMZoom catalog contains 816 detections and is $>95\%$ complete for compact sources with mass $>50~ M_{\odot}$, whereas the robust version of the catalog contains 285 detections and is $>95\%$ complete for sources with mass $>80~ M_{\odot}$. The effective radii of sources range between $0.04-0.4$ pc, and are likely not tracing individual cores, but rather groups of cores. A full description of the methods used to generate the CMZoom compact continuum source catalog can be found in \cite{Hatchfield_2020}.

\subsection{Herschel Column Density Map}
\label{herschel}

We use the column density map presented in \cite{Battersby_2025a} to quantify the local column density at the location of our ACES catalog sources. This column density map has an angular resolution of 36\arcsec ($\sim$1.4 pc at a distance of 8.178 kpc) and was generated from Herschel observations from the Hi-GAL Survey \citep[Herschel Infrared
Galactic Plane Survey;][]{Molinari_2010, Molinari_2016}. Further details on the methods used to derive this column density map can be found in \cite{Battersby_2025a}.

\section{Methods} 
\label{sec:methods}

In this section, we describe our method for extracting compact sources in the ACES continuum data and performing quality assurance on our detections. In Section \ref{subsec:dendro_methods}, we detail our procedure for creating a highly complete version of the ACES catalog and in Section \ref{subsec:tgif} we describe our method for fitting 2D Gaussians to each source. In Section \ref{subsec:manual_removal}, we describe our process for manually removing non-compact sources and noisy detections from the catalog using Zooniverse (\url{https://www.zooniverse.org/}), an online, volunteer-driven platform for large-scale data analysis. We outline our procedure for cross-matching our sources with other existing catalogs in Section \ref{subsec:mw_assoc}, and on flagging foreground and extended free-free emission contaminated detections in Sections  \ref{subsec:foreground} and \ref{subsec:ff_contam}. We describe the cuts we use to produce the final catalog in Section \ref{subsec:final_cat}. In Table \ref{tab:cat_cuts}, we summarize the steps used to obtain the final catalog using the methods described in this section. We provide a truncated version of the final ACES catalog in Tables \ref{tab:cat} and \ref{tab:cat2}. 

\subsection{Source Extraction using Dendrograms}
\label{subsec:dendro_methods}

In order to extract sources for the ACES compact continuum source catalog, we generate dendrograms using the \verb|astrodendro|\footnote{Detailed information on the \texttt{astrodendro} Python package can be found at \url{http://www.dendrograms.org/}.} Python package. Dendrograms are tree-like diagrams used to represent the hierarchical organization contained within a given data set. The ``branches'' of the dendrogram are structures that contain subsequent sub-structures, while the leaves are structures with no additional substructure. Used on our data, the leaves correspond to the most dense and compact structures, while the branches would be the less dense medium surrounding the leaves. For our catalog, we only include the dendrogram leaves, as these represent the densest structures in our observations.

We choose to use \texttt{astrodendro} since this algorithm does not assume an underlying morphology for our compact sources, which may be well-approximated by a two-dimensional Gaussian or instead exhibit more irregular border morphologies. 

The dendrogram we generate is determined using four parameters: 

\begin{enumerate}
    \item Minimum structure value, $f_{\text{min\_val}}$: the lowest allowed pixel value for a given structure.
    \item Minimum peak value, $f_{\text{min\_peak}}$: the lowest allowed peak pixel value for a given structure.
    \item Minimum significance, $\delta$: the threshold for how high a structure pixel value must be in comparison to a nearby structure to be considered an independent structure.
    \item Minimum number of pixels, $n_{\text{pix}}$: the minimum number of pixels required for a structure to be considered an independent structure in the dendrogram.
\end{enumerate}

We scale $f_{\text{min\_val}}$ and $\delta$ by factors of the global RMS noise value $\sigma_{\text{global}} = 100 ~ \mu$Jy/beam (see Section \ref{subsec:cont_data}). In an effort to prioritize completeness in the initial extraction of dendrogram leaves, we set $f_{\text{min\_val}} = 3\sigma_{\text{global}}$, $f_{\text{min\_peak}} = 4\sigma_{\text{global}}$, and $\delta = 0.5\sigma_{\text{global}}$. These values were chosen by manually confirming the highest thresholds we could set without removing certain ``benchmark'' sources; detections that are known to be real but are faint in the ACES mosaic. We describe these sources and why they are good ``benchmark'' candidates in Appendix \ref{dendro_justify}.

To capture all possible peaks in emission, we choose a low $n_{\text{pix}} = 7$, which is approximately a quarter of the number of pixels contained in the Gaussian area of our synthesized beam\footnote{$A_{\text{beam}} = \frac{b_{\text{maj}}b_{\text{min}}\pi}{4 \ln{2}}$, where $b_{\text{maj}}$ and $b_{\text{min}}$ are the FWHM of the major and minor axes of the beam.}. We use this lenient $n_{\text{pix}}$ value since point sources are best described by an elliptical 2D Gaussian, where the flux distribution extends beyond the FWHM limits of the beam. As a result using the full area of the Gaussian beam for dendrogram source extraction risks the removal of real sources. Additionally, some sources are located within negative `bowls', imaging artifacts located near bright sources of emission, and this artificially reduces their `true' peak intensity value. We discuss how we correct the peak intensity value for sources in Section \ref{subsec:tgif}.

Beyond these initial dendrogram parameters, we require each source to meet additional thresholds to be included in the catalog. First, we remove sources near the edges of the map where the observations are more impacted by noise due to the applied primary beam correction. To do this, we mask the catalog by requiring the leaf pixels to be at least 20 pixels from the edge of the ACES mosaic, which corresponds to $\sim 4$ times the beam width. 

Since the RMS noise varies across the ACES field, we use the ACES continuum noise map described in Section \ref{subsec:alma_obs} to determine a local RMS noise estimate ($\sigma_{\text{local}}$) for each leaf. To do this, we calculate the mean value of the ACES continuum noise map for pixels within the leaf contour. 

We also exclude catalog detections that are located in regions that are heavily impacted by imaging artifacts: the Sgr B2 region and the circumnuclear disk region near Sgr A*. Since there exists high-resolution observations of Sgr B2 at 3 mm \citep{Ginsburg_2018} which can be used as a supplement to this catalog, the excluded Sgr B2 region corresponds to the footprint of those observations. We exclude the noise-affected region around Sgr A* using a circular region centered at ($l,b$) = (359.94\degree, -0.046\degree) with a radius of 1.5\arcmin. The total area that our compact source catalog covers is shown within the black, dashed contours in Figure \ref{fig:aces_mosiac}, which also depicts the boundaries of our excluded regions.  

After this initial dendrogram extraction phase, which is built to be inclusive to all significant peaks in emission, we detect a total of 11523 sources. This is not the same as the number of sources in our final catalog, but is used as a highly complete base-level catalog that we refine further in Sections \ref{subsec:tgif} and \ref{subsec:manual_removal}. 

\subsection{Source fitting using TGIF}
\label{subsec:tgif}
Since our goal is to extract the most compact sources for our catalog, we use the 12m ACES continuum mosaic, which filters out larger scale structures from extended emission sources. However, there are some especially bright regions that exhibit imaging artifacts due to the lack of ``short-spacing'' observations \citep{Ginsburg_2026}. As a result, some real sources are situated in ``negative bowls'' while others are located on an enhanced background. These artifacts can result in the inaccurate determination of the peak intensity for a given dendrogram leaf. To correct for this, we use the \verb|TGIF| Python package \citep{yoo_2024}, which uses the \verb|lmfit| package to fit a 2D Gaussian ellipse to compact sources in astronomical observations. Using \verb|TGIF|, we are able to make a local background noise estimate using the median pixel value within an annulus surrounding the source, which we choose to scale with the effective radius of the dendrogram leaf, calculated as $R_{\text{leaf}} \equiv (N_{\text{pix}} A_{\text{pix}} / \pi)^{1/2}$, where $N_{\text{pix}}$ is the total number of pixels in the leaf, and $A_{\text{pix}}$ is the physical area of each pixel in pc$^{2}$. We use an annulus with an inner radius of $2\times R_{\text{leaf}}$ and an outer radius of $3\times R_{\text{leaf}}$ to calculate the local background surrounding the source. The only other parameter that we input is the ``fitting size'' which determines the side length of the cutout image that is used when making the fit. We set this parameter to be equal to the effective diameter of the dendrogram source to limit the fit to the profile of the compact source. We note that the local background is calculated directly from an annulus region of the original data, and not from the smaller cutout image. Ultimately, this fitting procedure results in a corrected peak intensity and integrated flux, as well as an estimate of the major and minor axes of the source. We compare the dendrogram peak intensity to the TGIF corrected peak intensity for each source in Appendix \ref{tgif_dendro_comp}. The TGIF fit is also complementary to the dendrogram source extraction procedure since it identifies the central `core' component of the leaf, which may include background material not associated with the source within its boundary.

\begin{figure*}
\begin{centering}
\epsscale{1.0}
\plotone{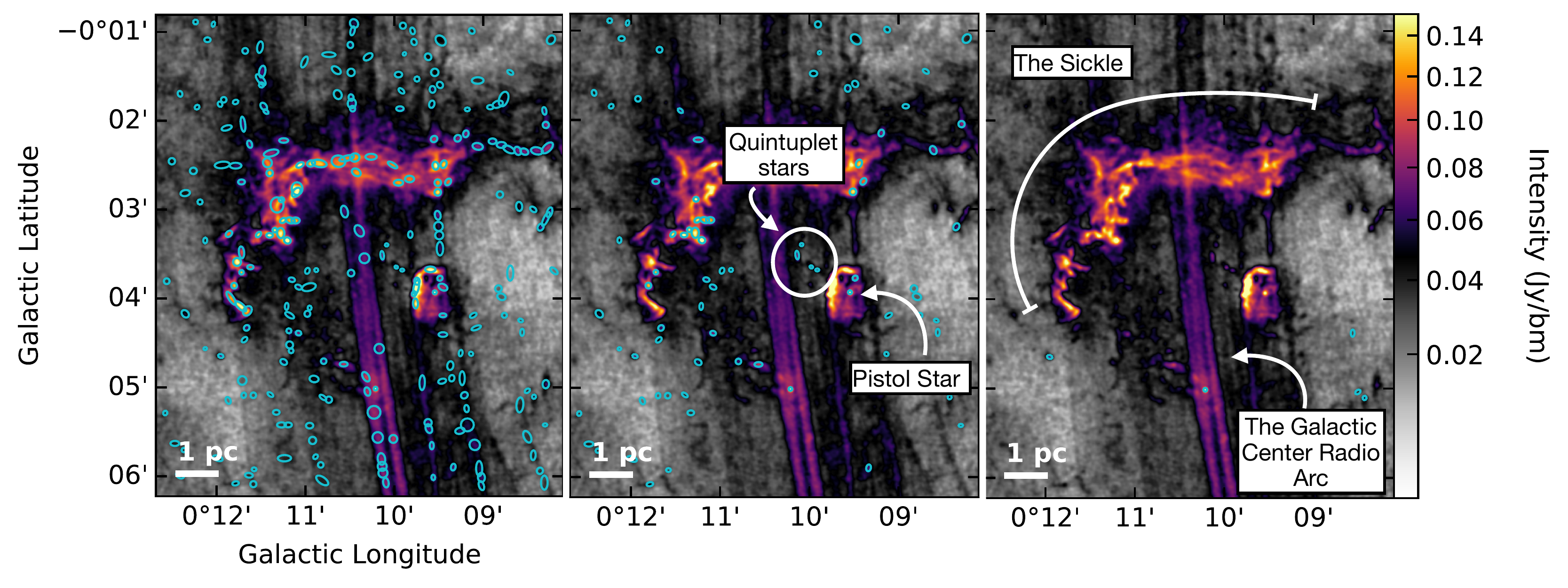}
\caption{A visual representation of ACES catalog sources at various steps in the pruning process. The image shown in each panel is a zoom-in of the ACES 3 mm continuum mosaic towards a group of non-thermal filaments collectively known as the the ``Galactic Center Radio Arc'' \citep{Yusef-Zadeh_1986} and an extended H\textsc{ii} region known as ``the Sickle'' \citep[e.g.,][]{Yusef-Zadeh_1987}, which is likely ionized by stars in the nearby Quintuplet cluster. In all three panels, ACES detections are shown with cyan ellipses. The left panel shows the sources before manual source removal, the middle panel shows the ``full'' catalog sources, and the right panel shows ``filtered'' catalog sources. We note that there are some known compact sources such as the Pistol Star and some evolved stars from the Quintuplet cluster that are a part of the ``full'' catalog, but not the ``filtered'' catalog due to the strict thresholds we set for extended free-free contamination in Section \ref{subsec:ff_contam}.}  
\label{fig:before_after}
\end{centering}
\end{figure*}

\subsection{Manual Source Removal using Zooniverse}
\label{subsec:manual_removal}
After implementing the automated procedures explained in Sections \ref{subsec:dendro_methods} and \ref{subsec:tgif}, we require that the TGIF-corrected peak intensity for each source be $>3\sigma_{\text{global}}$ and $>2\sigma_{\text{local}}$. These parameters result in a source catalog containing 6214 detections.

While the interferometric process of creating the ACES 12m continuum mosaic removes large-scale emission, any compact ``knots'' of the larger-scale structure would still be present in our maps. For example, the bright central portions of the Galactic Center Radio Arc non-thermal filaments are still visible in our data since they are extended only in one dimension, meaning that they are not filtered out by the interferometer. Although the source of this emission is extended in nature, the peaks in emission are still detected as significant sources in our dendrogram-based extraction procedure, as shown in cyan contours in the left panel of Figure \ref{fig:before_after}. 

\begin{figure}
\begin{centering}
\epsscale{0.8}
\plotone{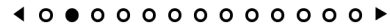}
\caption{Page 2 of the Zooniverse tutorial developed for classifying the compactness of ACES detections. The complete figure set (13 images) of the Zooniverse tutorial is available in the online journal.}
\label{fig:zootorial}
\end{centering}
\end{figure}

Due to the non-homogeneity of these sources and their properties, we have determined that a non-automated, by-eye approach is required to remove source detections originating from larger-scale emission. To accomplish this task, we used the Zooniverse (\url{https://www.zooniverse.org/}) interface to implement this procedure and retrieve classification data for each source. We required that each source was examined and categorized based on its ``compactness'' by 15 different members of the ACES team. In total, 49 ACES team members participated in this classification procedure. We provided all participants with the same information and instructions for classifying sources, including an in-built Zooniverse tutorial that was used to inform volunteers on the general features of the data as well as examples of compact, extended, and noisy detections. An example of one page from the tutorial is shown in Figure \ref{fig:zootorial}. 

\noprint{\figsetstart}
\noprint{\figsetnum{3}}
\noprint{\figsettitle{Zooniverse Tutorial}}

\figsetgrpstart
\figsetgrpnum{3.1}
\figsetgrptitle{Page 1}
\figsetplot{step1.png}
\figsetgrpnote{Page 1 of the Zooniverse tutorial developed for classifying the compactness of ACES detections.}
\figsetgrpend

\figsetgrpstart
\figsetgrpnum{3.2}
\figsetgrptitle{Page 2}
\figsetplot{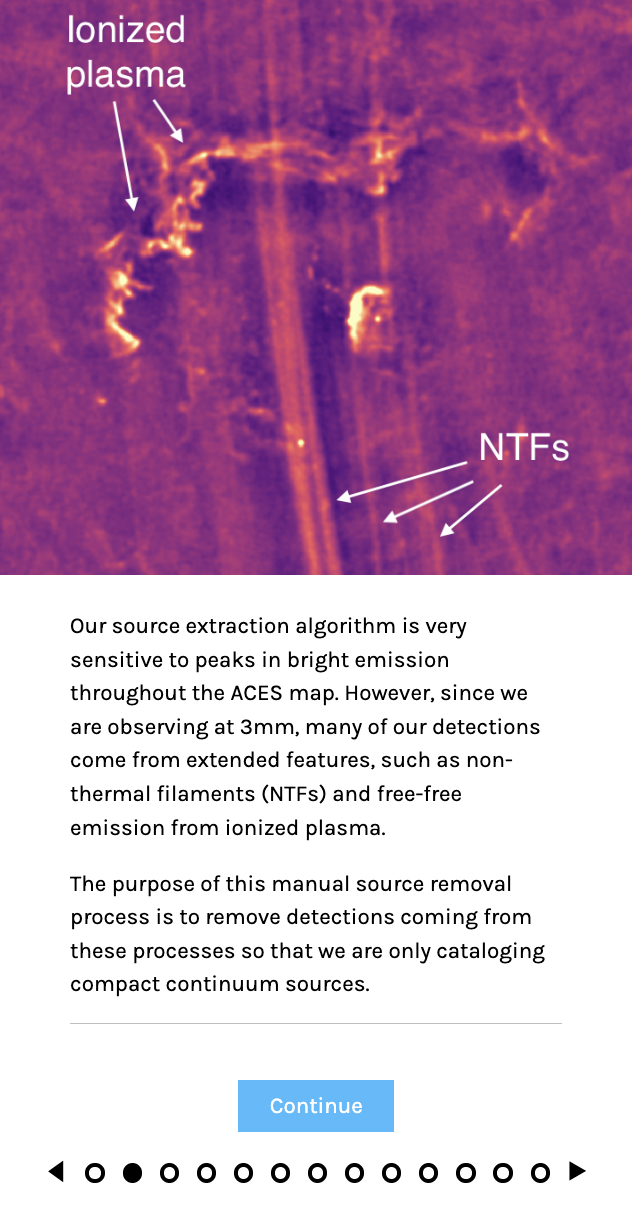}
\figsetgrpnote{Page 2 of the Zooniverse tutorial developed for classifying the compactness of ACES detections.}
\figsetgrpend

\figsetend

\begin{figure*}
\begin{centering}
\includegraphics[width=0.85\textwidth]{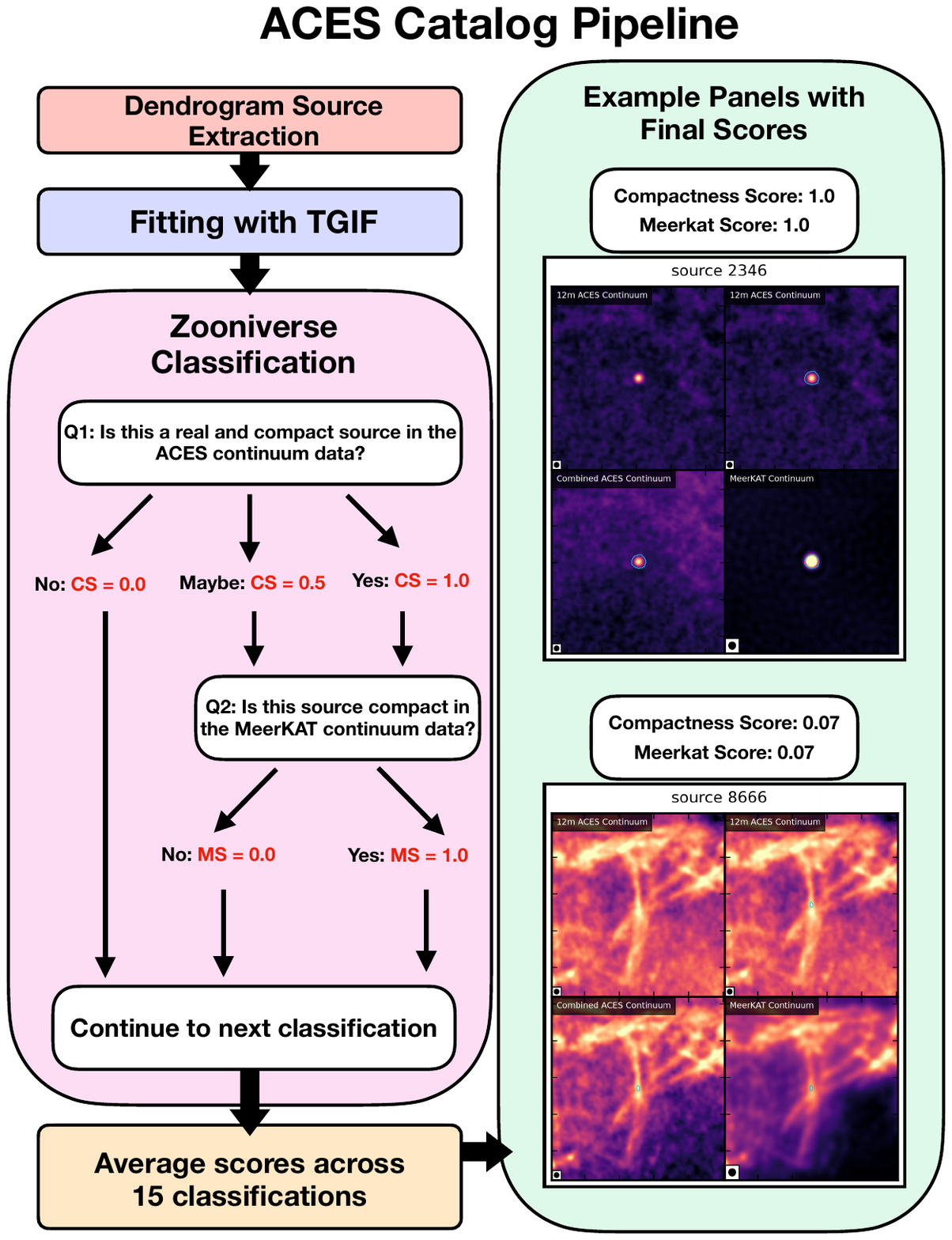}

\caption{The ACES catalog pipeline, including a decision tree for the Zooniverse source classification procedure outlining the questions presented to volunteers, the possible responses, and the Compactness Score (CS) and MeerKAT Score (MS) corresponding to each response. In the final section of the flow chart (green box), we present two examples of the 4-panel figures used for classification in the Zooniverse user interface. Each panel is centered on the coordinates of the source being classified. The top panels show the 12m ACES continuum mosaic image, the bottom left panel shows the ACES mosaic combined with single-dish GBT observations, and the bottom right panel shows the MeerKAT 1.28 GHz continuum image. The top right and bottom panels include the dendrogram contour in cyan. The synthesized beam for each map is shown in the bottom left corner of each panel. The unique source ID is given at the top of the figure, corresponding to the ``ACES ID'' column of the catalog (Table \ref{tab:cat}).}
\label{fig:zoo_flow}
\end{centering}
\end{figure*}

For each source, we provide a 4-panel figure containing images centered on the source being categorized (see images in Figure \ref{fig:zoo_flow}). The different panels show the cataloged source in the ACES 12m continuum image (top panels, with and without the dendrogram leaf contour for clarity), the ACES single dish combined continuum image, and the MeerKAT continuum image (both with the dendrogram leaf contour).
Using this set of images, each volunteer must determine if a source is a true, compact source detection. The ACES single dish combined continuum image provides context on extended emission, and the MeerKAT continuum image provides context on the nature of the source emission, since non-thermal filaments and free-free emission are detected at 1.28 GHz.

In Figure \ref{fig:zoo_flow}, we provide a flow chart detailing the decision tree for classifying sources.
We ask each user the following question: ``Is this source compact in the ACES continuum data?'', which the user can then respond with ``Yes'', ``No'', or ``Maybe''. If the user answers this question with ``Yes'' or ``Maybe'', they will be asked: ``Is this source compact in the MeerKAT continuum data''.
Each response has a numeric score associated with it: a response of ``Yes'' is scored a 1, a response of ``No'' is scored a 0, and a response of ``Maybe'' is scored a 0.5. For each source, we calculate the mean score for each question using a jackknife resampling technique, which first calculates the mean for each sub-sample of the data where one value has been removed, and then takes the mean of the sub-sample means. We refer to the score for question 1 as the ``compactness score'' (CS) and the score for question 2 as the ``MeerKAT score'' (MS). 

\subsection{Corresponding detections at different wavelengths}
\label{subsec:mw_assoc}
There are several known compact emission mechanisms at 3 mm, such as thermal emission from pre-/protostellar cores and the dusty envelopes of evolved stars, free-free emission from H\textsc{ii} regions, and non-thermal synchrotron radiation from the radio lobes of external galaxies. To constrain which physical process is producing emission for the sources in our catalog, we cross-reference our detections with objects observed at millimeter, radio, near-infrared (NIR), and X-ray wavelengths. Additionally, we note if any sources have a corresponding detection in the Set of Identifications, Measurements, and Bibiliography for Astronomical Data (SIMBAD) database \citep{Wenger_2000}. We provide a summary of the cross-matched detections in Table \ref{tab:class}. To perform this cross-matching, we use the CDS XMatch service from the astroquery Python package \citep{Ginsburg_2019}.

At millimeter wavelengths, we cross-match our catalog with the ``complete'' version of the 1.3 mm CMZoom compact continuum source catalog. We consider an ACES detection to have a match in the CMZoom catalog if the ACES dendrogram leaf has any amount of overlap with the CMZoom dendrogram leaf. We report the leaf ID of the CMZoom structure in the ``CMZoom ID'' column of the catalog (Table \ref{tab:cat}). If an ACES leaf overlaps with multiple CMZoom leaf structures, we report the ID corresponding to the CMZoom leaf with the greatest overlap.

To determine if an ACES source is associated with a compact radio detection, we leverage previously cataloged compact H\textsc{ii} regions in the CMZ, as well as the MeerKAT Score (MS) that we measured in Section \ref{subsec:manual_removal}. A recent catalog of candidate H\textsc{ii} regions was published in \cite{Lu_2019b}, using \textit{C}-band (5.56 GHz) observations from the Very Large Array (VLA) with an angular resolution of $\sim 1$\arcsec. We cross-reference  our sources with the detections in \cite{Lu_2019b}, and consider a source to be associated with a radio detection if its central coordinate is within 3\arcsec~ of the central coordinate of one of their reported candidate H\textsc{ii} regions.

In addition to this, we can estimate the total number of ACES sources that have an analogous compact detection in the 1.28 GHz MeerKAT continuum data by setting a threshold on the MS. For this analysis, we consider a source to be associated with a compact MeerKAT detection if it has a mean MS $\geq 0.8$. We choose this strict threshold for MS since we are interested only in sources that are robustly associated with compact MeerKAT emission.

We are specifically interested in NIR detections corresponding to evolved stars such as asymptotic giant branch (AGB) and Red Supergiant (RSG) stars, which have absolute magnitudes < 0. We cross-reference with NIR detections from the Two Micron All Sky Survey point source catalog \citep{Cutri_2003} and and the Spitzer Galactic Legacy Infrared Midplane Survey Extraordinaire (GLIMPSE) source catalog (I + II + 3D) \citep{IPAC_2009}.  A source is considered associated with a NIR detection if it is within 3\arcsec~ of a 2MASS catalog point source with an observed $K_{s}$-band magnitude < 10 or a GLIMPSE catalog point source with an observed $4.5~\mu$m magnitude < 9. We choose these thresholds since the CMZ has a distance modulus of 14.5, a $K_{s}$-band extinction of $A_{\text{K}_{\text{s}}} \sim 2.5$, and a 4.5$\mu$m extinction of $A_{[4.5\mu]} \sim 1 $ \citep{Fritz_2011}.

We consider a source to be associated with an X-ray source if it is within 3\arcsec~of a \textit{Chandra} detection reported in \cite{Muno_2009} or in the \textit{Chandra} source catalog \citep{Evans_2024} (CSC 2.1). The \cite{Muno_2009} catalog contains 9017 X-ray detections from \textit{Chandra} observations that cover a $2\degree \times 0.\degree8$ field around the Galactic center. The CSC contains 407,806 X-ray detections in 15,533 observations made with \textit{Chandra}. 

To obtain literature-based identifications and cross-identifiers, we also cross-reference our sources with objects in the SIMBAD astronomical database, requiring that the ACES detection be within 3\arcsec of a given SIMBAD object. We consider any spatially correlated SIMBAD detection to be a match, and do not restrict cross-identifications by object type or size.

If we do not find any associated detections using the methods described in this Section, we flag the source as ``unassociated''. The results of our multi-wavelength cross-matching are shown in Table \ref{tab:class}.

\begin{deluxetable*}{cccccccc}
\tabletypesize{\small}
\tablecaption{Step-by-step summary of catalog construction, further detailed in Section \ref{sec:methods}. \label{tab:cat_cuts}}
\startdata
\tablehead{
  \colhead{Phase in catalog creation} & 
  \colhead{Total N$_{\text{source}}$} &
  }
Initial Dendrogram Extraction & 11523 \\
Applying 2$\sigma_{\text{local}}$ and 3$\sigma_{\text{global}}$ cut & 6214 (Zooniverse sample) \\
Applying a stricter 4$\sigma_{\text{global}}$ cut & 2355 \\
Excluding sources with CS $< 0.5$ & 1735 (Full Catalog)\\ 
Flagging foreground detections and potential sources of extended free-free contamination  & 567 (Filtered Catalog)
\enddata
\end{deluxetable*}

\subsection{Known foreground detections}
\label{subsec:foreground}

Along the line-of-sight towards the CMZ, there is a dense molecular cloud known as the ``Pillar'', which was suggested to be in the foreground by \cite{Lu_2019b}, since one of the masers they detected within the cloud is spatially coincident with a bright source that is at a heliocentric distance of $\sim740$ pc, based on  parallax measurements made with the \textit{Gaia} satellite \citep{Gaia_collab_2018}. 

The ``Pillar'' cloud can be seen as a ``leaf'' structure in the dendrogram decomposition of the Herschel column density map in  \cite{Battersby_2025b}. This structure corresponds to ID 20 from their dendrogram catalog. All ACES detections located within the bounds of this dendrogram structure have been flagged as potential foreground detections in the ``FG'' column of the catalog (Tables \ref{tab:cat2}).

In \cite{Battersby_2020}, the authors listed five observed regions from the SMA CMZoom survey that were potentially in the foreground of the CMZ. Only three of these observed regions are located within the ACES field of view: G0.393-0.034, G0.212-0.001, and G359.137+0.031. The G359.137+0.031 observation is towards the ``Pillar'' cloud. As described in \cite{Battersby_2020}, there was no conclusive evidence that the other two regions, G0.393-0.034 and G0.212-0.001, were located outside of the CMZ, and as a result we do not flag these sources as ``FG'' detections.

There is a known stellar cluster ``DB00-58'' located at $l=359.994\degree,~b=0.156\degree$ that was identified using \textit{Chandra} X-ray observations in \cite{Law_2004}. This cluster of young stars has a measured $A_v$ that is inconsistent with what is expected in the Galactic center, and so it is considered a foreground cluster. However, after cross-matching with our catalog, we find that there are no ACES detections within the cluster core radius of $\sim22\arcsec$.

In \cite{Gramze2025}, the authors identify a foreground star-forming filament (G0.342+0.024) using JWST observations. This filament contains two protostellar cores that were observed in ACES 3 mm continuum, with protostellar outflows that were found using the ACES SiO J$=2 \xrightarrow{}1$ spectral line data. These cores, dubbed C1 and C2 have Galactic coordinates ($0.3315\degree, 0.0223\degree$) and ($0.3366\degree, 0.0239\degree$), respectively. We find that C1 and C2 correspond to the compact sources with IDs 18897 and 19018 in our ACES catalog, which we flag as foreground sources.

\subsection{Contamination from extended free-free emission}
\label{subsec:ff_contam}

Although the Compactness Score (CS) is a useful metric for removing sources that are associated with extended emission from non-thermal filaments, the manual classification procedure is less robust at removing sources of extended free-free emission.

In Figure \ref{fig:before_after}, we show a zoom-in from the ACES 3 mm continuum map. In this selected region, there are two well-known extended emission features: the ``Galactic Center Radio Arc'' \citep{Yusef-Zadeh_1986}, which is composed of vertical, non-thermal filaments, and the ``Sickle'', which is emission from gas that has been ionized by the nearby Quintuplet cluster \citep{Yusef-Zadeh_1987}. As shown in the left and middle panels, removing sources with a CS $< 0.5$ removes most of the detections from the ``Sickle'' and almost all of the detections from the NTFs, however there are some ``knots'' in ionized emission that are still considered compact detections. To account for this, we develop a method for flagging sources that are potentially associated with extended free-free emission.

To avoid flagging known compact sources, we first assert that all sources with associated compact radio detections are not considered ``free-free contaminated''. Additionally, we consider ACES sources with corresponding detections in the ``complete'' CMZoom catalog and an estimated 1 mm -- 3 mm spectral index ($\alpha_{\text{ACES-CMZoom}} > 2$) to be robustly not free-free contaminated. For the spectral index calculation, we first convolve the ACES and SMA mosaics to a common resolution of 3.2\arcsec, the typical spatial resolution of the CMZoom continuum observations. Then we take the integrated flux from within the ACES dendrogram leaf at both 1 mm ($S_{\text{CMZoom}}$) and 3 mm ($S_{\text{ACES}}$) and input them into the following relation: 

\begin{equation}
\alpha_{\text{ACES-CMZoom}} = \frac{\log (S_{\text{ACES}}/S_{\text{CMZoom}})}{\log\left(97.21 ~\text{GHz}/226\text{~GHz}\right)}
\end{equation}

We impose the $\alpha_{\text{ACES-CMZoom}} > 2$ threshold because free-free emission can still be detected at 1 mm, so some detections present in both catalogs may still be from extended free-free emission. 

The catalogs we use to determine if sources have NIR and X-ray associations have a much higher number density on the plane of the sky towards the Galactic center ($> 1000$ sources deg$^{-2}$) than our radio and millimeter catalogs. As a result, the cross-matching between these catalogs and the ACES catalog are more likely to be coincidental along the line-of-sight, and as a result we do not exclude them from being flagged as potential extended free-free emission contaminants.

To flag sources that are potentially part of extended free-free emission, we measure the spectral index using the MeerKAT and ACES continuum mosaics. To do this, we use a spectral index map generated from these datasets. To create this data product, we convolved the single-dish combined ACES and MeerKAT continuum mosaics to a common beam size of 4\arcsec. From here, we compute the spectral index map using the following relationship:  

\begin{equation}
\alpha_{\text{MeerKAT-ACES}} = \frac{\log (S_{\text{MeerKAT}}/S_{\text{ACES}})}{\log\left(1.284 ~\text{GHz}/97.21 \text{~GHz}\right)}
\end{equation}

where $S_{\text{MeerKAT}}/S_{\text{ACES}}$ is the ratio of the continuum mosaics. The MeerKAT-ACES spectral index map can be accessed at Zenodo\dataset[DOI: 10.5281/zenodo.21414882]{https://doi.org/10.5281/zenodo.21414882}. For each ACES detection, we report the mean $\alpha_{\text{MeerKAT-ACES}}$ value from within the corresponding dendrogram leaf contour. We flag all sources with a flat or negative spectral index (mean $\alpha_{\text{MeerKAT-ACES}}<0.5$) as potential contamination from extended free-free emission.

The final, rightmost panel in Figure \ref{fig:before_after} shows only the ACES detections with a CS $\geq 0.5$ that are not flagged as potential free-free emission contaminants. We emphasize that these are strict criteria and that some of these flagged detections are compact sources that happen to be co-spatial with bright MeerKAT emission, such as for the recently discovered MUBLO object \cite{Ginsburg_2024}, which is located within 5\arcmin~ of Sgr A*, corresponding to a projected distance of $\sim12$ pc on the plane of the sky (assuming a GC distance of 8178 pc). We also note that this strict criterion may flag sources of non-thermal synchrotron radiation such as background active galactic nuclei or Galactic compact radio sources.

\subsection{The Final ACES Compact Continuum Source Catalog}
\label{subsec:final_cat}

Depending on the specific scientific-use case, the ACES catalog of 6214 sources can be pruned based on thresholds imposed for $\sigma_{\text{global}}$, $\sigma_{\text{local}}$, CS, and MS. For the catalog we present in this paper, we attempt to limit the number of false positive detections by imposing a more selective signal-to-noise threshold, requiring sources to have a TGIF-corrected peak intensity $>4\sigma_{\text{global}}$. This results in a catalog of 2355 sources. 

Since we are concerned with only compact detections, we require that sources have a mean CS $\geq0.5$. Sources that do not meet this criterion are considered to be either low-significance detections or are part of more extended structures. After removing these, we are left with 1735 compact continuum source detections remaining in the ACES catalog, approximately 28\% of the original 6214 sources. 

We consider this to be the ``full'' version of our catalog, which is highly inclusive to all compact sources above our detection limit. Of these 1735 sources, there are 1130 that are considered to be free-free contaminated and 58 that are flagged as potential foreground sources. The remaining 567 sources are considered ``high-confidence'' compact detections, with negligible contamination from extended emission sources. These sources make up the ``filtered'' ACES catalog, which is the catalog we report measurements for throughout this paper. We compare the ``full'' and ``filtered'' ACES catalogs in Appendix \ref{aces_compare}. 

\begin{figure*}
\begin{centering}
\plotone{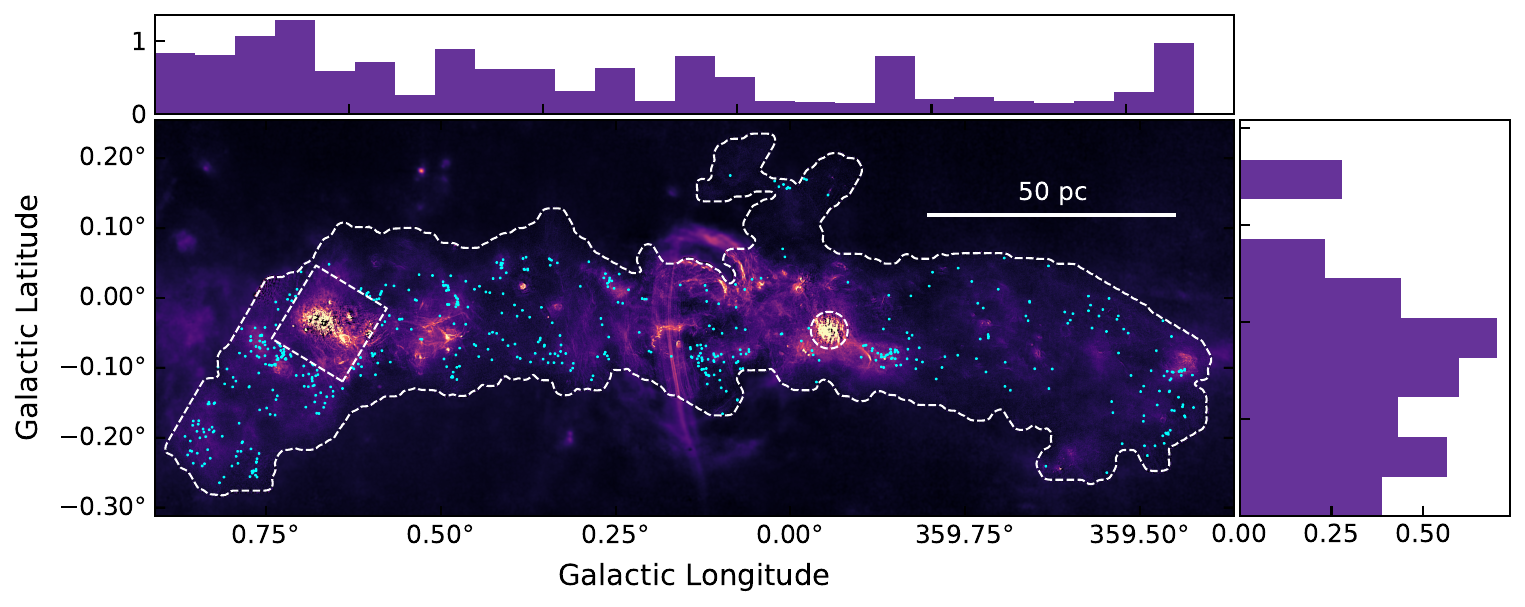}
\caption{The spatial distribution of source detections from the filtered version of the ACES compact continuum source catalog (cyan points) overlaid on 12m ACES 3 mm continuum mosaic combined with single-dish GBT observations. The white, dashed contours show the area of the CMZ that the ACES catalog covers. Sources found within the rectangular and circular regions are excluded from the catalog due to the impact of interferometric imaging artifacts.  The marginal histograms depict the number density of sources per square arcminute along the Galactic longitude and latitude axes.}
\label{lb_plot}
\end{centering}
\end{figure*}

\begin{splitdeluxetable*}{ccccccccccBccccccccc}
\tabletypesize{\scriptsize}
\tablecaption{A truncated version of the ``filtered'' ACES compact source catalog for 10 sources. This table includes the source identification number (ACES ID), position in Galactic coordinates ($l,b$), integrated flux ($F$), peak intensity ($I_{\text{peak}}$), FWHM major ($\theta_{\text{maj}}$) and minor ($\theta_{\text{min}}$) axes of its fitted ellipse, position angle with respect to the Galactic plane ($\theta_{\text{PA}}$), and the effective radius ($R_{\text{source}}$). We also include columns reporting the statistical uncertainties measured for  $\theta_{\text{maj}}$, $\theta_{\text{min}}$, $\theta_{\text{PA}}$, and $R_{\text{source}}$. The uncertainties estimated for $I_{\rm{peak}}$ and $F$, as described in Section \ref{subsec:phys_prop}, are also reported in this table. \label{tab:cat}}
\tablehead{
\colhead{ACES ID} & \colhead{$l$} & \colhead{$b$} & \colhead{$F$} & \colhead{$F$ unc.} &
\colhead{$I_{\text{peak}}$ } &
\colhead{$I_{\text{peak}}$ unc. } & 
\colhead{$I_{\text{dendro}}$ } & 
\colhead{$\sigma_{\text{local}}$} & \colhead{$\theta_{\text{maj}}$} & \colhead{$\theta_{\text{maj}}$ unc.} & \colhead{$\theta_{\text{min}}$} & \colhead{$\theta_{\text{min}}$ unc.} & \colhead{$\theta_{\text{PA}}$} & \colhead{$\theta_{\text{PA}}$ unc.} & \colhead{ $R_{\text{source}}$} & \colhead{ $R_{\text{source}}$ unc.} \\
\colhead{ } & \colhead{(deg)} & \colhead{(deg)} & \colhead{(Jy)} & \colhead{(Jy)} &
\colhead{(Jy beam$^{-1}$)} & 
\colhead{(Jy beam$^{-1}$)} & 
\colhead{(Jy beam$^{-1}$)} & 
\colhead{(Jy beam$^{-1}$)} & 
\colhead{(\arcsec)} & \colhead{(\arcsec)} & \colhead{(\arcsec)} & \colhead{(\arcsec)} & \colhead{(\degree)} & \colhead{(\degree)} & \colhead{(pc)} & \colhead{(pc)}}
\startdata
1903 & 0.8313 & -0.1884 & 7.08e-04 & 1.86e-03 & 0.000401 & 1.67e-04 & 4.70e-04 & 1.67e-04 & 2.86 & 0.10 & 2.80 & 0.10 & 143.97 & 6.60e+01 & 0.06 & 7.07e-02 \\
1965 & 0.8271 & -0.1848 & 2.25e-03 & 5.16e-03 & 0.000422 & 1.48e-04 & 4.25e-04 & 1.48e-04 & 5.44 & 0.42 & 4.45 & 0.28 & 143.69 & 1.97e+01 & 0.10 & 2.43e-01 \\
2028 & 0.8549 & -0.1822 & 1.91e-03 & 4.84e-03 & 0.000539 & 2.17e-04 & 6.24e-04 & 2.17e-04 & 4.66 & 0.03 & 3.45 & 0.01 & 139.17 & 6.85e-01 & 0.08 & 1.57e-02 \\
2044 & 0.8470 & -0.1805 & 2.22e-03 & 4.99e-03 & 0.000476 & 1.65e-04 & 5.31e-04 & 1.66e-04 & 5.81 & 0.40 & 3.65 & 0.16 & 192.50 & 3.19e+00 & 0.09 & 1.87e-01 \\
2055 & 0.7885 & -0.1791 & 3.07e-03 & 3.37e-03 & 0.000857 & 1.43e-04 & 9.89e-04 & 1.41e-04 & 4.99 & 0.21 & 3.26 & 0.10 & 134.60 & 2.91e+00 & 0.08 & 1.05e-01 \\
2101 & 359.4951 & -0.1765 & 2.77e-03 & 4.28e-03 & 0.000469 & 1.14e-04 & 5.24e-04 & 1.14e-04 & 5.77 & 0.18 & 4.64 & 0.12 & 267.24 & 4.00e+00 & 0.10 & 1.07e-01 \\
2116 & 0.8540 & -0.1760 & 1.83e-03 & 4.23e-03 & 0.000422 & 1.54e-04 & 4.32e-04 & 1.54e-04 & 5.49 & 0.23 & 3.59 & 0.08 & 191.15 & 1.84e+00 & 0.09 & 1.05e-01 \\
2120 & 0.8463 & -0.1755 & 3.37e-03 & 7.13e-03 & 0.000555 & 1.79e-04 & 6.41e-04 & 1.77e-04 & 8.06 & 0.70 & 3.42 & 0.11 & 140.15 & 2.45e+00 & 0.10 & 2.44e-01 \\
2243 & 359.4354 & -0.1703 & 2.56e-03 & 5.49e-03 & 0.000481 & 1.63e-04 & 6.22e-04 & 1.63e-04 & 5.83 & 0.14 & 4.15 & 0.07 & 265.27 & 1.71e+00 & 0.10 & 7.35e-02 \\
2286 & 359.4943 & -0.1683 & 1.16e-03 & 2.08e-03 & 0.000474 & 1.31e-04 & 6.81e-04 & 1.31e-04 & 3.53 & 0.16 & 3.15 & 0.14 & 220.18 & 9.63e+00 & 0.07 & 1.08e-01 \\
\enddata
\tablecomments{Both the ``full'' catalog and the ``filtered'' catalog (presented in Tables \ref{tab:cat} and \ref{tab:cat2}) are published in their entirety in the machine-readable format. }
\end{splitdeluxetable*}

\begin{splitdeluxetable*}{cccccccBccccccccccccc}
\tabletypesize{\scriptsize}
\tablecaption{A truncated version of the ``filtered'' ACES compact source catalog for 10 sources (continued). This table includes the source identification number (ACES ID), the mean CS and MS scores, the Herschel column density (N(H$_{2}$)) at the location of the source, columns flagging potential foreground (FG) and oversized (OS) sources, and the associated source matches from various catalogs outlined in Section \ref{subsec:mw_assoc}. \label{tab:cat2}}
\tablehead{
\colhead{ACES ID} & \colhead{Mean CS} & \colhead{Mean MS} & \colhead{N(H$_2$)} & \colhead{FG} &
\colhead{OS} & \colhead{CMZoom ID} & \colhead{2MASS ID} & \colhead{GLIMPSE ID} & \colhead{Lu+2019 ID} & \colhead{CSC ID} & \colhead{Muno+2009 ID} & \colhead{SIMBAD ID}\\
\colhead{ } & \colhead{} & \colhead{} & \colhead{(cm$^{-2}$)} & \colhead{} &
\colhead{} & \colhead{} & \colhead{} & \colhead{} & \colhead{} & \colhead{} & \colhead{}}
\startdata
1903 & 0.73 & 0.00 & 1.18e+23 & False & False & -1 & -- & 'G000.8315-00.1879' & -- & -- & -- & -- \\
1965 & 0.67 & 0.00 & 1.03e+23 & False & False & -1 & -- & 'G000.8266-00.1852' & -- & -- & -- & -- \\
2028 & 0.53 & 0.00 & 7.65e+22 & False & False & -1 & -- & -- & -- & -- & -- & -- \\
2044 & 0.70 & 0.00 & 9.60e+22 & False & False & -1 & -- & 'G000.8468-00.1803' & -- & -- & -- & --\\
2055 & 0.80 & 0.00 & 7.13e+22 & False & False & -1 & -- & 'G000.7889-00.1792' & -- & -- & -- & --\\
2101 & 0.93 & 0.00 & 5.16e+22 & False & False & 500 & -- & -- & -- & -- & -- & --\\
2116 & 0.50 & 0.00 & 7.85e+22 & False & False & -1 & '17481954-2817493' & 'G000.8544-00.1761' & -- & -- & -- & --\\
2120 & 0.67 & 0.00 & 8.62e+22 & False & True & -1 & -- & -- & -- & -- & -- & --\\
2243 & 0.87 & 0.00 & 2.30e+22 & False & False & -1 & -- & 'G359.4359-00.1706' & -- & -- & -- & --\\
2286 & 0.93 & 0.93 & 8.40e+22 & False & False & -1 & '17450415-2927216' & 'G359.4938-00.1685' & -- & -- & -- & --\\
\enddata
\tablecomments{Both the ``full'' catalog and the ``filtered'' catalog (presented in Tables \ref{tab:cat} and \ref{tab:cat2}) are published in their entirety in the machine-readable format. }
\end{splitdeluxetable*}

\section{Results} 
\label{sec:results}

\begin{figure*}
\begin{centering}
\plottwo{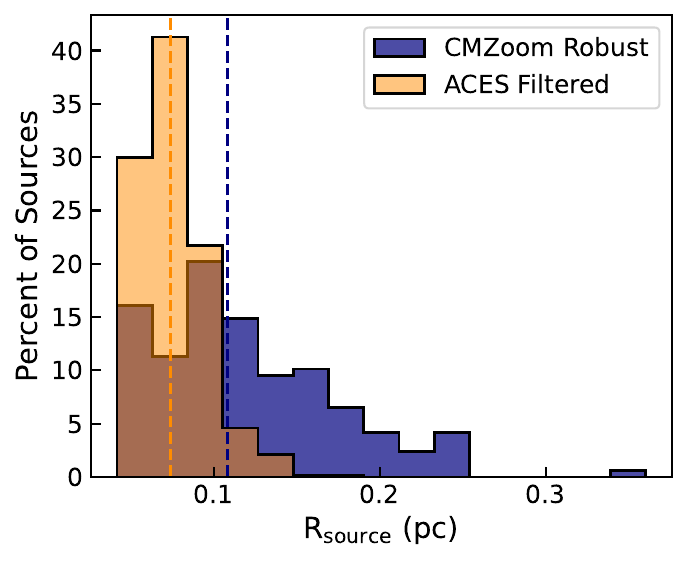}{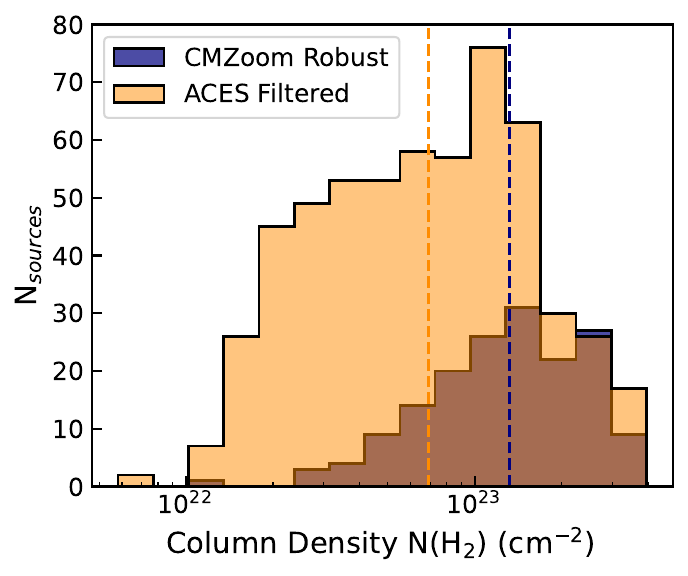}
\caption{Comparing the radius (left) and Herschel column density (right) distributions for detections in the filtered ACES (orange) and robust CMZoom (blue) compact continuum source catalogs. For ACES detections, the effective radius ($R_{\text{source}}$) is calculated from the geometric mean of the FWHM axes of the fitted TGIF ellipses and the CMZoom $R_{\text{leaf}}$ is calculated from the total area of its corresponding dendrogram leaf: $R_{\text{leaf}} = \sqrt{A_{\text{leaf}}/\pi}$. For both the ACES and CMZoom sources, the column density is averaged over the area of the leaf from the Herschel column density map. The dashed lines show the median values of each distribution.}
\label{fig:rad_nh2_hist}
\end{centering}
\end{figure*}

\subsection{Physical Properties}
\label{subsec:phys_prop}

We report on the physical properties for compact continuum sources in the ACES catalog (Tables \ref{tab:cat} and \ref{tab:cat2}). Both the ``full'' catalog and the ``filtered'' catalog are published in their entirety in the machine-readable format. For each source, we list its identification number (ACES ID), position in Galactic coordinates ($l,b$), integrated flux ($F$), TGIF-corrected peak intensity $I_{\text{peak}}$, FWHM major ($\theta_{\text{major}}$) and minor ($\theta_{\text{minor}}$) axes of its fitted ellipse, effective radius ($R_{\text{source}}$), CS and MS scores and its Herschel column density (N(H$_{2}$)). The coordinates of each source correspond to the position of peak intensity. We report the measured uncertainty for $I_{\rm{peak}}$ as the value of the ACES continuum noise map from section \ref{subsec:cont_data} at the position of the source peak intensity. We also provide statistical uncertainties for the measured quantities associated with the TGIF Gaussian fits.

The integrated flux of each source is calculated using 

\begin{equation} \label{eq:flux}
    F = \frac{I_{\text{peak}} 2\pi  \sigma_{\text{major}}\sigma_{\text{minor}}}{\Omega_{\text{beam}}},
\end{equation}

where $\sigma_{\text{major}}$ and $\sigma_{\text{minor}}$ are the major and minor axes of the fitted 2D Gaussian ellipse from TGIF, and $\Omega_{\text{beam}}$ is the solid angle area of the beam. We measure the uncertainty in the integrated flux by progagating the error of the measured properties in Equation \ref{eq:flux}, which we report in the catalog (see Table \ref{tab:cat}.)

We calculate the circularized angular size of each source by taking the geometric mean of the FWHM major and minor axes of its fitted ellipse.

\begin{equation}
    \theta_{\text{circ}} = \sqrt{\theta_{\text{major}}\theta_{\text{minor}}},
\end{equation}

Using this, we can estimate the linear size of the source in parsecs, and then divide this by 2 to get its effective radius $R_{\text{source}}$, which we report in the catalog (Table \ref{tab:cat}). In Figure \ref{fig:rad_nh2_hist}, we see that source radii range from 0.04 pc -- 0.18 pc, with a median $R_{\rm{source}}$ = 0.07 pc. The uncertainty for $R_{\rm{source}}$ reported in Table \ref{tab:cat} is estimated by propagating the statistical uncertainty of the major and minor axes of the fit while also incorporating a heliocentric distance error of $\pm150$ pc, since the distances to these sources within the radius of the CMZ is not well constrained. The column density reported in Table \ref{tab:cat2} is the Herschel column density value at the location of the peak pixel value for the source.

From the initial dendrogram extraction, we report the original estimate of the source peak intensity $I_{\rm{dendro}}$ and the local noise estimate $\sigma_{\rm{local}}$ that we used in Sections \ref{subsec:dendro_methods} and \ref{subsec:manual_removal} since these values were used to remove detections from noise. We do not report an effective radius or flux measurement from the dendrogram source extraction. This is because the dendrogram parameter $n_{\text{pix}}$ can be as small as $0.25\times$ the total number of pixels in the beam, leading to some sources having non-physical sizes that are smaller than the synthesized beam, causing underestimated integrated flux values. Additionally, the dendrogram flux measurement does not take the local background of the source into account.  

\begin{deluxetable*}{cccccccc}
\tabletypesize{\small}
\tablecaption{A multiwavelength summary of coincident detections at the cataloged source positions for both the filtered and full versions of the ACES catalog.\label{tab:class}}
\tablehead{
  \colhead{Catalog} & 
  \colhead{Total N$_{\text{source}}$} &
  \colhead{Millimeter} & \colhead{Radio} & 
  \colhead{NIR} & 
  \colhead{X-ray} &
  \colhead{SIMBAD Object} & 
  \colhead{Unassociated}
  }
\startdata
Filtered & 567 & 147 & 93 & 50 & 30 & 141 & 278 \\
Full & 1735 &  174 & 94 & 155  & 118 & 360 & 1137 \\
\enddata
\end{deluxetable*}

\subsection{Spatial Distribution of ACES Continuum Sources}
\label{subsec:spat_dist}

We show the spatial distribution of ACES compact continuum detections in Figure \ref{lb_plot}, with cyan points representing the source positions overlaid on the ACES single-dish combined continuum mosaic. The marginal histograms show the 1D source number density distributions in square arcminutes across Galactic longitude and latitude. We find that the spatial distribution of sources is asymmetric, with the higher source densities located at positive longitudes.  We note that the number density of sources between $l= 0.6\degree - 0.75\degree$ would likely be much higher if detections from Sgr B2 were included in this catalog. The source number density peaks at a Galactic longitude of $l\sim$ 0.7\degree. In Galactic latitude, the peak is located at $b\sim$ 0.2\degree, however these high-latitude sources may be previously unidentified foreground interlopers. If so, the ``true'' peak in source density for the CMZ is located at a Galactic latitude of $b\sim$ -0.05\degree, which would only have a higher density of sources if we included detections from the Sgr B2 region.

In the right panel of Figure \ref{fig:rad_nh2_hist}, we see that the Herschel column density for sources in the ACES catalog range from $\sim 6\times10^{21} - 4\times10^{23}$ cm$^{-2}$, with a median value of $\sim$ 6.9$\times10^{22}$ cm$^{-2}$. There are 359 sources located in regions where the Herschel column density is $< 10^{23}$ cm$^{-2}$, meaning that the majority of sources ($\sim$63$\%$) are located in lower column density regions of the CMZ. This is unexpected, since previous millimeter-wavelength studies of the CMZ focused only on the densest regions of molecular clouds where it is assumed that the vast majority of star forming cores are located \citep[e.g.][]{Hatchfield_2020}. 

\section{Discussion}
\label{sec:discussion}

\subsection{Nature of sources in the ACES catalog}
\label{subsec:nature_of_sources}

At a wavelength of 3 mm, the ACES continuum map is sensitive to thermal dust emission, free-free emission, and non-thermal synchrotron radiation. Using the manual classification procedure described in Section \ref{subsec:manual_removal}, we remove most of the sources associated with extended emission, such as non-thermal filaments associated with synchrotron radiation and ionized plasma from free-free emission, such as that observed in the Pistol Nebula (Figure \ref{fig:before_after}). In Section \ref{subsec:ff_contam}, we further refined our catalog by flagging sources that are potentially compact ``knots'' in extended free-free emission by calculating spectral indices, cross-referencing our catalog with known H\textsc{ii} regions, and by leveraging the MeerKAT score (MS) from our manual classification procedure.

After this rigorous pruning procedure, we are left with compact detections that are likely composed of embedded pre- and protostellar cores, compact H\textsc{ii} regions, evolved stars, and background galaxies. In Section \ref{subsec:mw_assoc}, we cross-reference our results with existing catalogs and the SIMBAD database to quantify the number of sources for which there are coincident detections at other observed wavelengths. We stress that these associated detections do not serve as an official ``classification'' for sources in our catalog, but are instead useful flags that can be used to better understand the nature of our sources.

In Table \ref{tab:class}, we provide a summary of the number of sources that have detections at the indicated wavelengths for both the ``filtered'' and ``full'' versions of our catalog.

\subsection{Comparing the ACES and CMZoom catalogs}
\label{sec:cmzoom_crossmatch}

It is useful to compare the continuum sources in the filtered ACES catalog with those identified in the CMZoom Survey catalog \citep{Hatchfield_2020}. The sources from the CMZoom catalog are largely associated with embedded star forming cores or groups of cores, with some possible contamination from free-free emission, typically from H$\textsc{ii}$ regions. In this section, we compare mostly to the ``robust'' version of the CMZoom catalog, but we include values from the ``complete'' catalog in parentheses.

\begin{figure}
\begin{centering}
\plotone{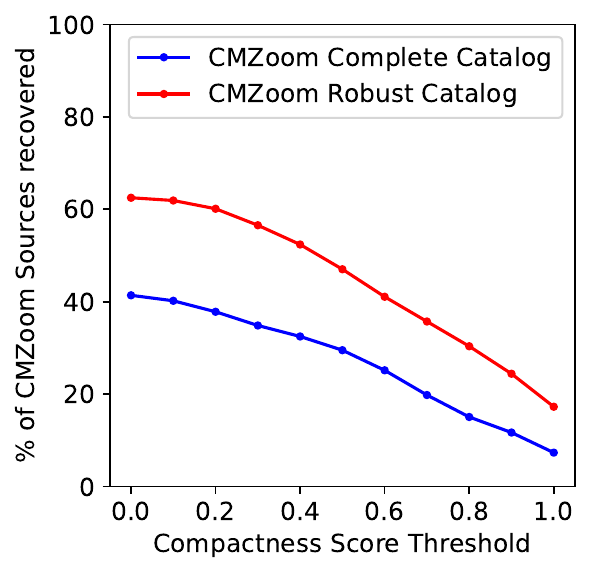}
\caption{The percentage of CMZoom detections recovered by the ``full'' ACES catalog as a function of our choice in the CS threshold. We report the percentage of sources recovered for both the robust (red) and complete (blue) versions of the CMZoom catalog \citep{Hatchfield_2020}. The percentage recovery is only calculated for the area where the ACES ALMA and CMZoom SMA observations overlap.} 
\label{fig:perc_recovered}
\end{centering}
\end{figure}

In total, the robust (complete) CMZoom catalog contains 285 (816) detections. Of these detections, only 168 (505) sources lie within the area covered by the ACES catalog (see the dashed-line boundary in Figure \ref{lb_plot}). This reduction in the number of overlapping detections is due to the removal of the highly populated Sgr B2 region from our analysis. We note that 268  ($\sim 47 \%$) of the sources from the ACES catalog are found within the observed area of the SMA CMZoom survey, meaning that the majority of our detections are located outside the dense, star forming regions observed with the SMA. In total, there are 90 (147) sources in the ACES catalog that overlap with 73 (129) CMZoom sources. The smaller number of CMZoom sources is likely due to the lower angular resolution of the SMA data, causing some sources that are resolved in ACES to be blended in CMZoom. 

For regions where the ACES and CMZoom survey overlap, the ACES catalog recovers $\sim$43.5\% ($\sim$25.5\%) of detections from the CMZoom catalog. Clearly, there are a considerable number of sources from the CMZoom catalog that we do not recover in the ACES catalog. There are a number of reasons for why this is the case. This discrepancy can be partially explained by the fact that our ``filtered'' catalog removes detections with CMZoom catalog counterparts if $\alpha_{\text{CMZoom}}<2$. If we instead consider our ``full'' catalog, we have matched detections with 79 (149) of the CMZoom sources, which results in only a minor improvement in the fraction of CMZoom sources recovered.

Another explanation is that the CS threshold used to generate the ``full'' ACES catalog is too conservative. In Figure \ref{fig:perc_recovered}, we examine how our choice in CS impacts the number of CMZoom sources recovered. We find that even at the least exclusive threshold of CS = 0, we reach a maximum percent recovery of $\sim 63\%$ ($\sim 41\%$), implying the possibility that the CMZoom catalog includes detections that may belong to more extended structures. Regardless, the manual source removal cannot fully explain the discrepancy. We also tried implementing a different cross-matching method, where ACES detections within 3\arcsec~ of a given CMZoom source are considered to be associated. Ultimately, we find that using a different cross-matching method does not change the number of CMZoom sources recovered.

We also consider the difference in mass sensitivity and completeness between the two catalogs. Although the continuum flux sensitivity is similar between the ALMA and SMA data, the mass sensitivity is different since the dust opacity per unit mass is higher at 1 mm than it is for 3 mm. Using our $f_{\text{min\_peak}} = 400~\mu$Jy/beam and assuming optically thin dust continuum emission, we can derive the mass sensitivity for the ACES catalog, 
\begin{equation}
    M_{\text{source}} = \frac{d^2 F R_{\text{gd}}}{\kappa_{\nu} B_{\nu}(T_{d})}
\label{mass_eq}
\end{equation}
where $d$ is the assumed distance to Galactic center of 8178 pc, $F$ is the integrated flux of the point source, $R_{\text{gd}}$ is the gas-to-dust ratio,  $\kappa_{\nu}$ is the dust opacity per unit mass, and $B_{\nu}(T_{d})$ is the Planck function at the local dust temperature $T_d$, which we take to be 20 K. We take the gas-to-dust ratio to be 100 and we use a dust opacity per unit mass of $\kappa_{\text{3mm}} = 0.18~$cm$^{2}$g$^{-1}$ based on the \cite{Ossenkopf_1994} model for moderately coagulated dust grains with thin ice mantles at gas densities $\geq 10^6$ cm$^{-3}$ and a coagulation timescale of 10$^5$ yr.

From this, we find that ACES catalog is sensitive to sources down to a mass of approximately 13.5 M$_{\odot}$. We then calculate masses for sources from the CMZoom catalog, using the same values for $T_{d}$, $d$, and $R_{\text{gd}}$, and  an assumed dust opacity per unit mass of $\kappa_{\text{1.3mm}} = 0.9~$cm$^{2}$g$^{-1}$. If we use the ACES ``full'' catalog (CS$\geq0.5$) and consider only CMZoom sources with masses $>13.5$ M$_{\odot}$ the percent of recovered CMZoom sources is still only 52.3\% (34.3\%). 

Using synthetic source extraction, \cite{Hatchfield_2020} determined that the robust version of the CMZoom catalog was complete to sources with masses $>80 M_{\odot}$. Since we removed sources from the ACES catalog using by-eye classifications (Section \ref{subsec:manual_removal}), it is difficult to accurately estimate what mass the ACES catalog is complete to. In light of this, we use the mass completeness limit of the CMZoom catalog and calculate the percent recovery after removing sources with $M < 80$ M$_{\odot}$ from the CMZoom catalog. By doing this, we recover 73\% (63\%) of the CMZoom sources with $M > 80$ M$_{\odot}$ when comparing to the ``full'' ACES catalog, and 67\% (58\%) when comparing to the ``filtered'' catalog. If we again consider the impact of the manual source removal and use the least exclusive threshold of CS = 0 (with only the 4830 sources with $I_{\text{peak}} > 4\sigma_{\text{global}}$ included), we find that we recover a maximum of 87\% (78\%), which means that there are still 13\% (27\%) of the CMZoom sources where no ACES counterpart is found. 

Another plausible explanation is that the ALMA and SMA data have unique imaging artifacts and varying local noise levels across their respective maps.  In Appendix \ref{aces_cmzoom_compare} we present side-by-side images of the ALMA and SMA continuum observations and compare the sources identified in the ACES and robust CMZoom catalogs. In multiple images, we find instances where the CMZoom catalog has detections that are not included in the ACES catalog due to being (a) not compact (b) located at the edge of the map where noise is higher, and (c) not being visible in the ACES data. This is usually because the CMZoom detection is located in a negative bowl in the ACES continuum image, or because it is itself a product of noise enhancement near negative bowls in the SMA data. We discuss this further in Appendix \ref{aces_cmzoom_compare}.

It is possible that some detections are only identified in either the ACES or CMZoom catalogs due to flux variability since these observations were obtained at different epochs. Millimeter-wavelength variability has been observed in protostellar clusters in the Galaxy, possibly due to episodic mass accretion events \citep[e.g.,][]{2015_Safron, Hunter_2017, Yang_2026}. However, these observations are very rare, so we do not consider this to be a major driver reducing the number of cross-matched detections.


The compact continuum sources identified in the robust CMZoom catalog have effective radii spanning 0.04 -- 0.4 pc, with a median value of $\sim0.1$ pc (Figure \ref{fig:rad_nh2_hist}). This range exceeds what is calculated for the ACES catalog, with source radii ranging between 0.04 -- 0.2 pc, and a median $R_{\text{source}}\sim0.07$ pc. Although this is in part due to the ALMA data having an angular resolution of $2.5\arcsec$ and the SMA data having a typical angular resolution of $3\arcsec$, it can also be attributed to the difference in how the source is defined. Although sources were extracted using \verb|astrodendro| for both the ACES and the CMZoom catalogs, the ACES source radii are not determined from the dendrogram leaf size as they are in CMZoom, but instead are calculated from the 2D elliptical Gaussian fit from TGIF. We also remove sources based on their relative `compactness' using the compactness score (CS) during the manual classification phase, a procedure that is not performed for the sources in the CMZoom catalog. 

We also compare the Herschel column density values for the ACES and CMZoom detections. In the right panel of Figure \ref{fig:rad_nh2_hist}, we see that CMZoom sources are located at column densities between $\sim 1 \times10^{22} - 2\times 10^{24}$ cm$^{-2}$, with a median N(H$_2$) $\sim 1\times 10^{23}$ cm$^{-2}$. We find that the ACES catalog has a similar number of sources as CMZoom down to N(H$_2$) $\sim 1\times 10^{23}$ cm$^{-2}$, but below this column density threshold there are significantly more detections in ACES than there are in CMZoom. This is likely due to the improved spatial coverage of the ACES survey in comparison to the CMZoom survey, which was limited in scope to just the densest molecular clouds in the CMZ.

\subsection{Sources at low column densities}
\label{subsec:lowdens_reasons}

In Section \ref{subsec:spat_dist}, we found that 359 ($\sim$63$\%$) of the compact continuum sources from the ACES catalog were found within relatively low column density regions of the CMZ (N(H$_2$) $< 1\times10^{23}$ cm$^{-2}$), outside of the major molecular clouds. This is an interesting result considering that the majority of high-resolution studies on embedded star formation in the CMZ focus exclusively on these incredibly dense regions \citep[e.g.][]{Ginsburg_2018, Hatchfield_2020, Walker_2018}. In light of this, we consider various reasons for why we have a large number of detections at N(H$_2$) $< 1\times10^{23}$ cm$^{-2}$.

Of the 359 compact continuum sources located at lower column densities, only 164 ($\sim$45.6\%) of them have an associated detection based on the catalog cross-matching we performed in Section \ref{subsec:mw_assoc}. The remaining 195 ($\sim$54.4\%) low-density detections have no known associations at these wavelengths.  Given this information, we consider it most likely that these sources are previously undetected ultra/hyper (UC/HC) compact H\textsc{ii} regions, which would likely be much brighter in the ACES continuum than the MeerKAT continuum due to their higher optical depth when compared to more evolved H\textsc{ii} regions.

Although we flag known foreground detections in the ``Pillar'' region and those identified in \citep{Gramze2025}, we also cannot currently exclude the possibility that other sources are located outside of the CMZ. However, we assume this does not account for the majority of the ``low-density'' sources in the catalog. Future analysis of core kinematics will be useful in confirming their membership in the CMZ.

Another explanation is that the Herschel column density map has an angular resolution of 36\arcsec, meaning that the smaller-scale fluctuations in column density are smoothed out in these data. So it is possible that some compact sources are located in denser substructures not resolved in the Herschel data.

If we assume that the majority of these detections are YSOs within the CMZ, it is possible that the cores initially form within the major cloud complexes, but then kinematically decouple from them at some point in their evolution. Recent magnetohydrodynamic (MHD) simulations from \cite{Tress_2025} find that stars and gas decouple quickly in the CMZ due to the short orbital timescales, resulting in O stars that become cyclically exposed and re-embedded. In the context of our observations, it is possible that we are tracing a population of YSOs that are no longer associated with their parent cloud. A thorough investigation of dynamical star formation in the CMZ using the ACES catalog is outside the scope of this work, and will be presented in a future paper (Battersby, et al. in prep.).

\subsection{Sources of Uncertainty}
\label{subsec:error}

There are a number of potential sources of uncertainty for the analysis presented in this paper. For the flux and size measurements, the largest source of error comes from the initial elliptical Gaussian fit made using TGIF. Although the majority of our compact detections are well described by this morphology, there are a number of cases where the underlying emission is not well represented by a 2D Gaussian ellipse, either due to interferometric artifacts or because of background emission. Typically, these sources are ``over-sized'' in nature since the fit could not converge to a central ``point'' source. We flag sources with a major axis FWHM $> 3 \times$ the beam width of 2.56\arcsec~ as being ``over-sized'' and determine that only $4.4\%$ of our filtered catalog and $3.7\%$ of our full catalog are affected. These sources are flagged in the `OS' column of the catalog (see Table \ref{tab:cat2}). If we exclude these sources from the analyses performed in this paper, we find that it does not significantly impact our primary results. However, we encourage future users of the catalog to be careful in interpreting measurements from these individual sources.

\section{Conclusions}
\label{sec:conclusions}

In this paper, we described our procedure for cataloging compact continuum sources in the ACES 3 mm continuum mosaic and presented their fundamental characteristics. We extracted sources using a dendrogram-based algorithm, and then used the Python package TGIF to fit each source with a 2D Gaussian and to estimate a corrected peak intensity based on its local background. We removed sources with a corrected peak intensity < 3$\sigma_{\text{global}}$ and 4$\sigma_{\text{local}}$, where $\sigma_{\text{global}}$ and $\sigma_{\text{local}}$ are noise estimates for the full ACES mosaic as well as the local noise estimate at the location of the source, respectively.

After this automated procedure, we used a manual source classification procedure implemented in the Zooniverse interface to help with removing sources associated with extended sources of non-thermal or free-free emission. The resulting output results in a mean compactness score (CS) and MeerKAT score (MS), which measure the relative ``compactness'' and association with compact radio detections for each source, depending on the by-eye classifications of users.

For the analysis in this paper, we present a science-ready version of the ACES catalog, requiring sources to have a mean CS > 0.5, a peak intensity > 4$\sigma_{\text{global}}$, and removing known foreground sources and sources that may be compact ``knots'' in otherwise extended ionized emission. We provide a summary of the ACES catalog used in this paper and related findings below:

\begin{enumerate}
    \item The filtered ACES compact continuum source catalog contains 567 detections spanning effective radii of 0.04 - 0.18 pc, with a median value of 0.07 pc. 
    \item There are 268 ($\sim47\%$) detections located in regions covered by both the ACES and CMZoom surveys. In this overlap region, the ACES catalog recovers 43.5\% of the sources from the robust CMZoom catalog.  
    \item We find that the majority of sources in the ACES catalog ($\sim$ 63$\%$) are located at N(H$_2$) $< 1 \times 10^{23}$ cm$^{-2}$, placing them outside the densest molecular cloud regions in the CMZ.  
    \item We cross-match ACES detections with catalogs at radio, millimeter, infrared, and X-ray wavelengths. In total, 45.6\% of cataloged sources have associations with detections at other wavelengths. The remaining 54.4\% of sources are potentially new detections, not previously identified in the external catalogs we cross-reference. 
\end{enumerate}

In future studies, we will use our catalog cross-referencing to thoroughly classify these continuum sources as H\textsc{ii} regions, dusty star forming regions or evolved stars based the wavelengths they are detected at. We will also use the abundance of high-resolution molecular line observations from ACES to determine the kinematics and chemistry of sources, properties that will be important for placing their membership in the CMZ. With this information, we will be able to more accurately constrain the star-forming potential of the CMZ and better understand the evolution and dynamics of protostars in this unique environment. 


\section{acknowledgements}

\noindent J.W. gratefully acknowledges funding from National Science Foundation under Award Nos. 2108938 and 2206510.
\\
C.B.  gratefully  acknowledges  funding  from  National  Science  Foundation  under  Award  Nos. 2108938, 2206510, 2414862, and CAREER 2145689, as well as from the National Aeronautics and Space Administration through the Astrophysics Data Analysis Program under Award ``3-D MC: Mapping Circumnuclear Molecular Clouds from X-ray to Radio,” Grant No. 80NSSC22K1125 as well as participation in the PRIMA project under Grant No. 80NSSC25K7944.
\\
MGSM thank the Spanish MCINN for funding support under grant PID2023-146667NB-I00 funded by MCIN/AEI/10.13039/501100011033.
MGSM acknowledges support from the NSF under grant CAREER 2142300.
\\
I.J-.S, L.C., and V.M.R acknowledge funding from grant PID2022-136814NB-I00 funded by the Spanish Ministry of Science, Innovation and Universities/State Agency of Research MICIU/AEI/ 10.13039/501100011033 and by “ERDF/EU”. I.J.-S. also acknowledges support from the ERC grant OPENS, GA No. 101125858, funded by the European Union.
\\
RSK acknowledges financial support from the ERC via Synergy Grant ``ECOGAL'' (project ID 855130) and from the German Excellence Strategy via the Heidelberg Cluster ``STRUCTURES'' (EXC 2181 - 390900948). In addition RSK is grateful for funding from the German Ministry for Economic Affairs and Climate Action in project ``MAINN'' (funding ID 50OO2206), and from DFG and ANR for project ``STARCLUSTERS'' (funding ID KL 1358/22-1).
\\
FNL gratefully acknowledges financial support from the Ramón y Cajal programme (RYC2023-044924-I) funded by MCIN/AEI/10.13039/501100011033 and FSE+, from grant PID2024-162148NA-I00, funded by MCIN/AEI/10.13039/501100011033 and the European Regional Development Fund (ERDF) ‘A way of making Europe’ , and from the Severo Ochoa grant CEX2021-001131-S, funded by MCIN/AEI/10.13039/501100011033.
\\
The project that gave rise to these results received the support of a fellowship from the ``la Caixa'' Foundation (ID 100010434). The
fellowship code is LCF/BQ/PR25/12110012.
\\
A.S.-M. acknowledges support from the RyC2021-032892-I grant funded by MCIN/AEI/10.13039/501100011033 and by the European Union `Next GenerationEU’/PRTR, as well as the program Unidad de Excelencia María de Maeztu CEX2020-001058-M, and support from the PID2023-146675NB (MCI-AEI-FEDER, UE).
\\
KMD acknowledges support from the European Research Council (ERC) Advanced Grant MOPPEX 833460.
\\
This research made use of astrodendro, a Python package to compute dendrograms of Astronomical data ~(\url{http://www.dendrograms.org/}).
\\
This publication uses data generated via the Zooniverse.org platform, development of which is funded by generous support, including a Global Impact Award from Google, and by a grant from the Alfred P. Sloan Foundation.
\\
The authors would like to thank Greg Troiani for helpful conversations regarding the style of Zooniverse projects and their data exports.
\\
The National Radio Astronomy Observatory is a facility of the National Science Foundation operated under cooperative agreement by Associated Universities, Inc.
\\
This paper makes use of the following ALMA data: ADS/JAO.ALMA\#2021.1.00172.L. ALMA is a partnership of ESO (representing its member states), NSF (USA) and NINS (Japan), together with NRC (Canada), MOST and ASIAA (Taiwan), and KASI (Republic of Korea), in cooperation with the Republic of Chile. The Joint ALMA Observatory is operated by ESO, AUI/NRAO and NAOJ.
\\

\facilities{ALMA}
\software{Astropy \citep{astropy:2013, astropy:2018, astropy:2022}, Astrodendro (\url{http://www.dendrograms.org/}), CASA \citep{CASA_2022}, Numpy \citep{2020NumPy-Array}, TGIF \citep{yoo_2024}}

\appendix

\section{``Benchmark'' sources used to determine initial dendrogram parameters}
\label{dendro_justify}

\begin{figure*}
\begin{centering}
\plottwo{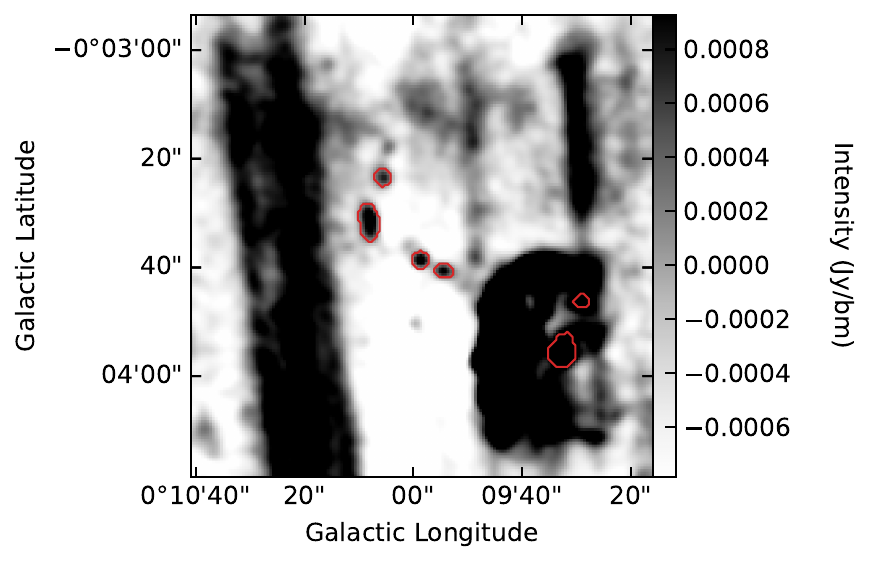}{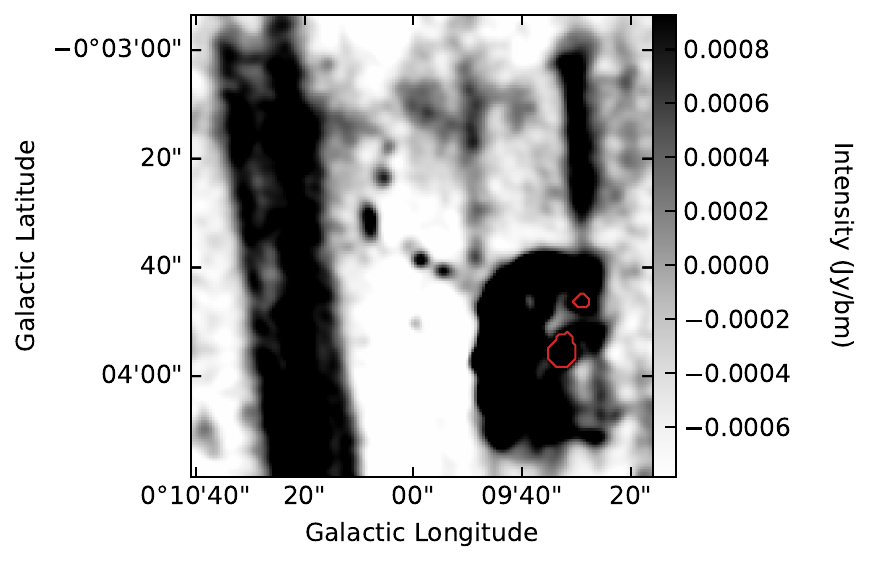}
\caption{A side-by-side comparison of the dendrogram extraction for evolved stars from the Quintuplet cluster, as seen in the ACES 3 mm continuum map. These evolved stars are located in a negative ``bowl'' artifact in the interferometric image. The left panel image shows the ACES dendrogram leaf mask contours for sources in the ``full'' catalog presented in this paper, with $f_\text{min\_val} = 3\sigma_{\text{global}}$. The right panel image shows the leaf mask contours for a dendrogram generated with $f_\text{min\_val} = 4\sigma_{\text{global}}$. }
\label{fig:benchmark}
\end{centering}
\end{figure*}

In Section \ref{subsec:dendro_methods}, we report on the input parameters used to perform the initial dendrogram source extraction. The chosen parameters were designed to create a highly complete catalog of sources that would then be refined based on the manual quality assurance procedure performed using the Zooniverse interface, as described in Section \ref{subsec:manual_removal}. As a result, we used certain previously identified, faint sources to inform our decision for how strict these dendrogram parameter choices could be. One such group of previously identified sources are the evolved stars near the Quintuplet cluster that were located within negative `bowl' imaging artifacts in the ACES data, which artificially reduced their peak brightness. As shown in Figure \ref{fig:benchmark}, the highest integral factor of $f_{\text{min\_val}}$ we could use for dendrogram source extraction was $3\sigma_{\text{global}}$. If we set the threshold any higher, we would fail to detect these real, compact sources. The remaining noise-based thresholds that we set in Sections \ref{subsec:dendro_methods} and \ref{subsec:tgif} were chosen using the same methodology.

\section{Comparing the TGIF corrected peak intensity to the initial dendrogram peak intensity}
\label{tgif_dendro_comp}

\begin{figure*}
\begin{centering}
\plottwo{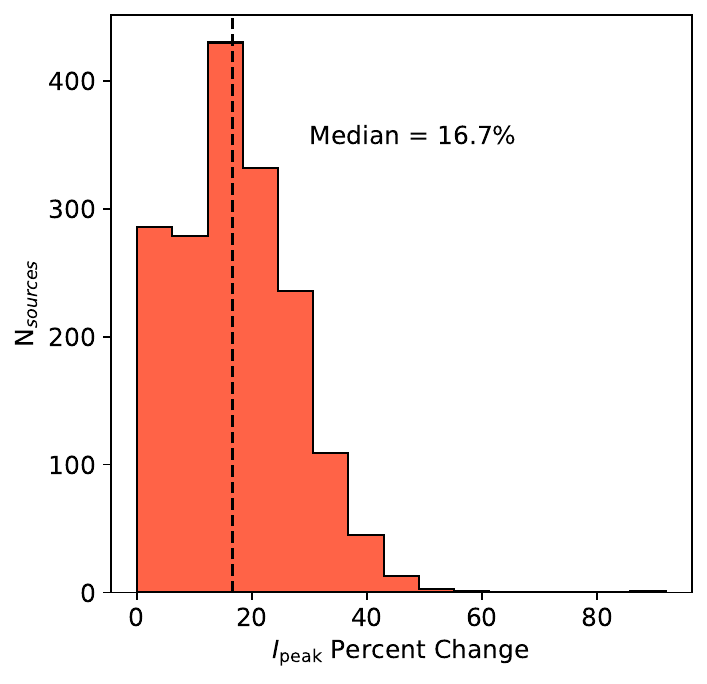}{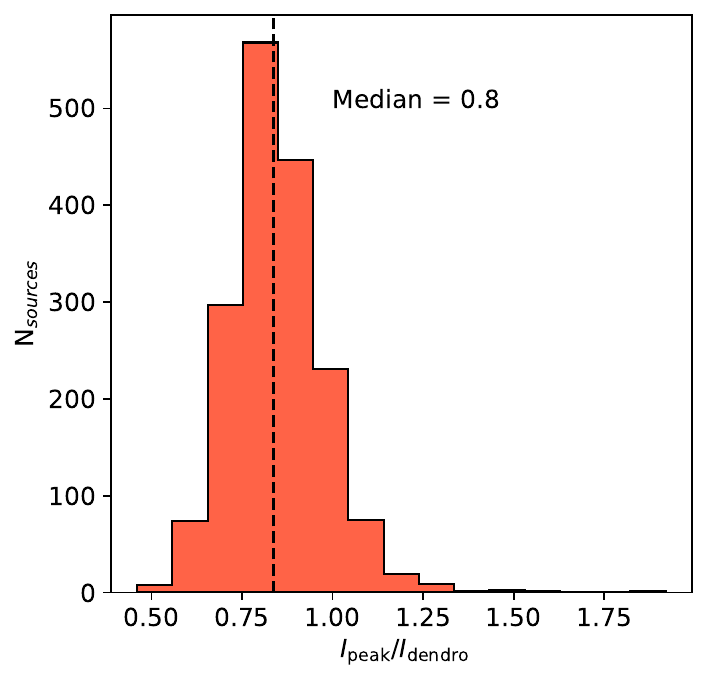}
\caption{Two histograms characterizing the difference in the initial peak intensity, $I_{\text{peak}}$, of ACES sources extracted by the dendrogram and the corrected peak intensity calculated using TGIF. Both histograms are using the ``full'' version of the catalog. The left panel is a histogram of the percent change in $I_{\text{peak}}$, and the right panel is a histogram of the $I_{\text{peak}}$ ratio. The dashed vertical lines indicate the median value for each distribution.}
\label{fig:peak_int_comp}
\end{centering}
\end{figure*}

In Section \ref{subsec:tgif}, we discuss how we use TGIF to fit each source extracted using astrodendro. This fit estimates a local background for each source, and produces a corrected peak intensity, $I_{\text{peak}}$. In Figure \ref{fig:peak_int_comp}, we compare the percent change of $I_{\text{peak}}$ as well as the $I_{\text{peak}}$ ratio between the dendrogram and TGIF outputs. We find that overall, the corrected $I_{\text{peak}}$ is typically somewhat lower than the original value, $I_{\rm{dendro}}$, with the median ratio $I_{\text{peak}}$/ $I_{\rm{dendro}}$ = 0.8. In general, the percent change in peak intensity is between 0--40\%, with a median percent change of 16.7\%.

\section{Comparing the Full and Filtered versions of the ACES catalog}
\label{aces_compare}

\begin{figure*}
\begin{centering}
\plottwo{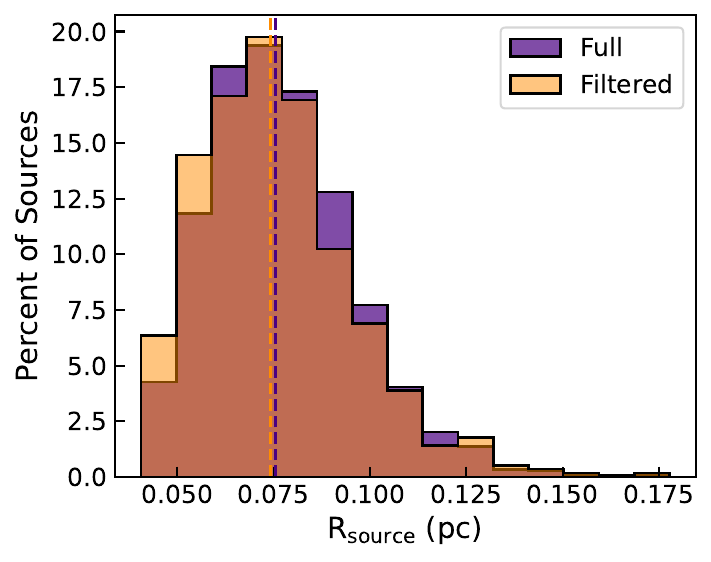}{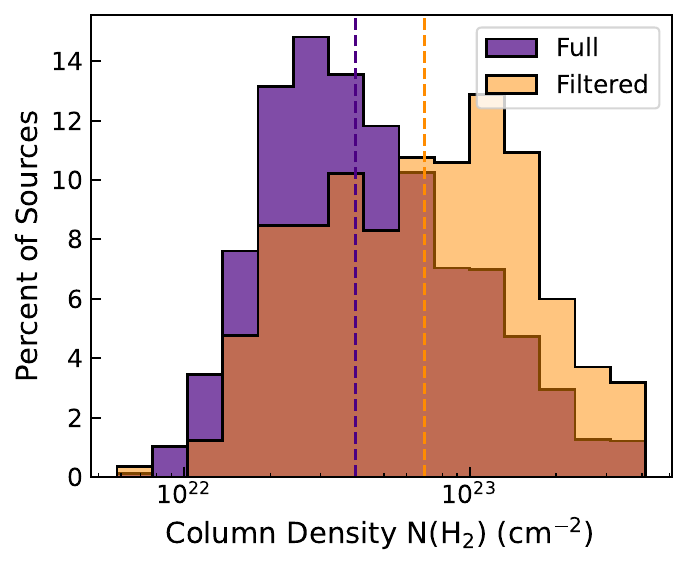}
\caption{Comparing the radius (left) and Herschel column density (right) distributions for detections in the filtered (orange) and full (purple) versions of the ACES compact continuum source catalog. The effective radius ($R_{\text{source}}$) is calculated from the geometric mean of the FWHM axes of the fitted TGIF ellipses. The column density is the Herschel column density map value at the location of the peak pixel of the ACES source. The dashed lines show the median values of each distribution.}
\label{fig:rad_nh2_hist_comp_rob}
\end{centering}
\end{figure*}

In Section \ref{subsec:final_cat}, we defined the ``full'' and ``filtered'' versions of the ACES catalog. For the analysis in this paper, we primarily use the ``filtered'' version of the catalog, however we considered it important to see how our results would change with the ``full'' catalog. In the left panel of Figure \ref{fig:rad_nh2_hist_comp_rob} we find that the effective radius distribution of sources changes by a negligible amount between catalog versions. In the right panel of Figure \ref{fig:rad_nh2_hist_comp_rob} there is a notable shift in the overall Herschel column density distribution and its median moves to lower densities for the ``full'' catalog. This is largely due to the high density of ``free-free contaminated'' sources located in the ACES continuum mosaic between $l \sim 359.6 \degree - 0.3 \degree$.

\section{Comparing the ACES and CMZoom catalogs}
\label{aces_cmzoom_compare}

\begin{figure*}
\begin{centering}
\epsscale{1.0}
\plottwo{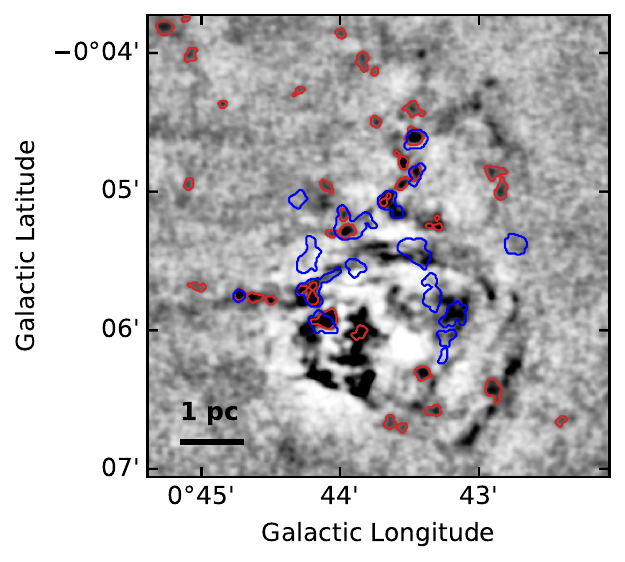}{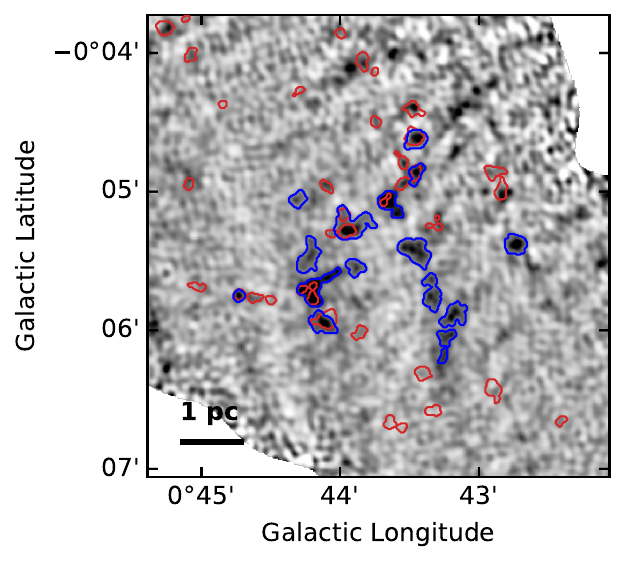}
\plottwo{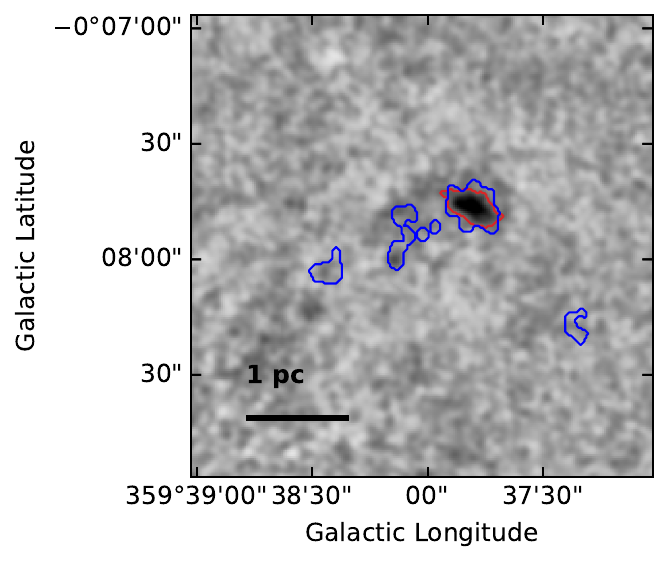}{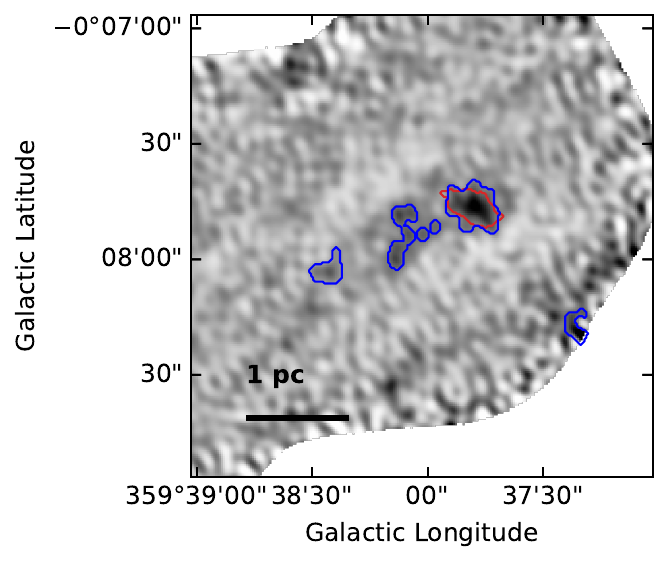}
\plottwo{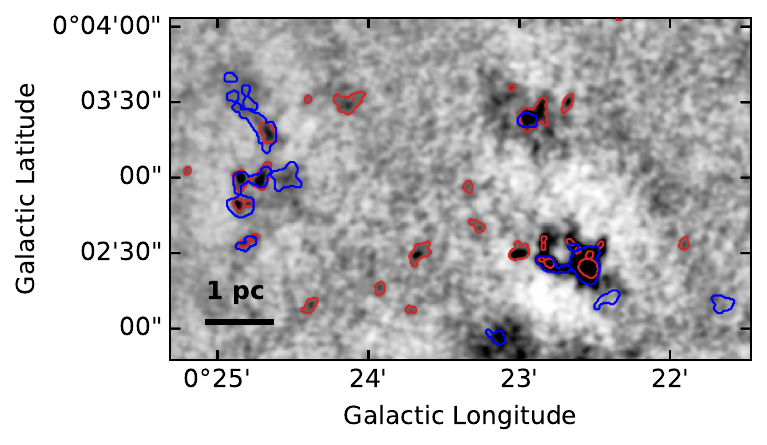}{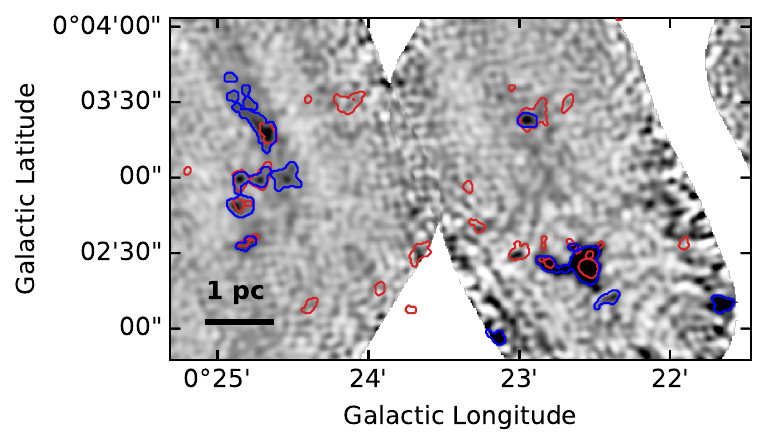}
\caption{Three side-by-side comparisons of 3 mm ALMA (left) and 1 mm SMA (right) observations. Red and blue contours show the leaf boundaries for the ACES catalog and CMZoom catalog sources, respectively. We use the full version of the ACES catalog and the robust version of the CMZoom catalog for these figures. A 1 pc scalebar is provided in the bottom left corner of each image.}
\label{fig:cmzoom_aces_compare1}
\end{centering}
\end{figure*}

As noted in Section \ref{sec:cmzoom_crossmatch}, we reported the somewhat low percentage recovery of CMZoom sources with the ACES catalog. Although this is partially due to our choice in pruning thresholds, it can also be due to differences in the source extraction technique and the interferometric nature of the data. Although the ALMA and SMA data are relatively similar in sensitivity and angular resolution, the distribution of interferometric noise and imaging artifacts varies between observations. Additionally, the CMZoom catalog has a lower limit mass sensitivity than the ACES catalog, so sources that are faint in the SMA data may be undetectable in the ALMA data. In Figure \ref{fig:cmzoom_aces_compare1}, we highlight a few regions where these factors play a role in why the ACES catalog fails to detect sources from the robust CMZoom catalog.

\bibliography{references}{}
\bibliographystyle{aasjournal}

\end{document}

%% file: authors.tex
\author[0009-0002-7459-4174]{Jennifer Wallace}
\affiliation{University of Connecticut, Department of Physics, 196A Auditorium Road Unit 3046, Storrs, CT 06269 USA}
\email{jennifer.2.wallace@uconn.edu}

\author[0000-0002-6073-9320]{Cara Battersby}
\affiliation{University of Connecticut, Department of Physics, 196A Auditorium Road Unit 3046, Storrs, CT 06269 USA}

\author[0000-0002-0533-8575]{Nazar Budaiev}
\affiliation{Department of Astronomy, University of Florida, PO Box 112055, Gainesville, FL 32611 USA}

\author[0000-0003-0946-4365]{H Perry Hatchfield}
\affiliation{University of Connecticut, Department of Physics, 196A Auditorium Road Unit 3046, Storrs, CT 06269 USA}

\author[0000-0003-0410-4504]{Ashley T. Barnes}
\affiliation{European Southern Observatory (ESO), Karl-Schwarzschild-Stra{\ss}e 2, 85748 Garching, Germany}

\author[0000-0003-2968-5333]{Taehwa Yoo}
\affiliation{Department of Astronomy, University of Florida, PO Box 112055, Gainesville, FL 32611 USA}

\author[0000-0002-3941-0360]{Miriam G. Santa-Maria}
\affiliation{Department of Astronomy, University of Florida, PO Box 112055, Gainesville, FL 32611 USA}
\affiliation{Instituto de Física Fundamental (CSIC), Calle Serrano 123, 28006 Madrid, Spain}

\author[0009-0004-6748-721X]{Antonio Daley} 
\affiliation{Department of Astronomy, University of Florida, PO Box 112055, Gainesville, FL 32611 USA}

\author[0000-0002-9279-4041]{Q. Daniel Wang}
\affiliation{Department of Astronomy, University of Massachusetts, Amherst, MA 01003, USA}

\author[0000-0002-1730-8832]{Anika Schmiedeke}
\affiliation{Green Bank Observatory, P.O. Box 2, Green Bank, WV 24944, USA}

\author[0009-0005-8382-0614]{Kaitlyn E. Sheriff} 
\affiliation{Department of Physics and Astronomy, University of Kansas, 1251 Wescoe Hall Drive, Lawrence, KS 66045, USA}

\author[0000-0003-3341-6144]{Jairo Armijos-Abenda\~no}
\affiliation{Observatorio Astron\'omico de Quito, Observatorio
Astron\'omico Nacional, Escuela Polit\'ecnica Nacional, 170403, Quito,
Ecuador}

\author[0000-0001-7330-8856]{Daniel Walker}
\affiliation{UK ALMA Regional Centre Node, Jodrell Bank Centre for Astrophysics, The University of Manchester, Manchester M13 9PL, UK}

\author[0000-0001-5933-2147]{Gwenllian Williams}
\affiliation{Department of Physics, Aberystwyth University, Ceredigion, Cymru, SY23 3BZ, UK}

\author[0000-0002-5776-9473]{Dani R. Lipman}
\affiliation{University of Connecticut, Department of Physics, 196A Auditorium Road Unit 3046, Storrs, CT 06269 USA}

\author[0000-0003-3390-4893]{Farideh Mazoochi}
\affiliation{Institute for Research in Fundamental Sciences (IPM), School of Astronomy, Tehran, Iran}

\author[0000-0002-4013-6469]{Natalie O. Butterfield}
\affiliation{National Radio Astronomy Observatory, 520 Edgemont Road, Charlottesville, VA 22903, USA}

\author[0000-0002-6379-7593]{Francisco Nogueras-Lara}
\affiliation{Instituto de Astrof\'{i}sica de Andaluc\'{i}a, CSIC, Glorieta de la Astronom\'ia s/n, 18008 Granada, Spain}

\author[0000-0002-4268-6499]{Yoshiaki Sofue}
\affiliation{Institute of Astronomy, The University of Tokyo, Mitaka, Tokyo, 181-0015, Japan}

\author[0000-0002-0560-3172]{Ralf S.\ Klessen}
\affiliation{Universit\"{a}t Heidelberg, Zentrum f\"{u}r Astronomie, Institut f\"{u}r Theoretische Astrophysik, Albert-Ueberle-Str.\ 2, 69120 Heidelberg, Germany}
\affiliation{Universit\"{a}t Heidelberg, Interdisziplin\"{a}res Zentrum f\"{u}r Wissenschaftliches Rechnen, Im Neuenheimer Feld 225, 69120 Heidelberg, Germany}

\author[0000-0001-6708-1317]{Simon C. O. Glover}
\affiliation{Universit\"{a}t Heidelberg, Zentrum f\"{u}r Astronomie, Institut f\"{u}r Theoretische Astrophysik, Albert-Ueberle-Str.\ 2, 69120 Heidelberg, Germany}

\author[0000-0003-4140-5138]{Katharina Immer}
\affiliation{European Southern Observatory (ESO), Karl-Schwarzschild-Stra{\ss}e 2, 85748 Garching, Germany}

\author[0000-0002-1313-429X]{Savannah R. Gramze}
\affiliation{Department of Astronomy, University of Florida, PO Box 112055, Gainesville, FL 32611 USA}

\author[0000-0003-0980-6871]{Katarzyna M. Dutkowska}
\affiliation{Leiden Observatory, Leiden University, P.O. Box 9513, 2300 RA Leiden, The Netherlands}

\author[0009-0004-0685-7678]{Rojita Buddhacharya}
\affiliation{Astrophysics Research Institute, Liverpool John Moores University, IC2, Liverpool Science Park, 146 Brownlow Hill, Liverpool L3 5RF, UK}
\affiliation{Center for Astrophysics | Harvard \& Smithsonian, 60 Garden Street, Cambridge, MA 02138, USA}

\author[0000-0001-8064-6394]{Laura Colzi}
\affiliation{Centro de Astrobiolog\'ia (CAB), CSIC-INTA, Ctra. de Ajalvir Km. 4, 28850, Torrej\'on de Ardoz, Madrid, Spain}

\author[0000-0002-1254-4174]{Ashley Lieber}
\affiliation{Department of Physics and Astronomy, University of Kansas, 1251 Wescoe Hall Drive, Lawrence, KS 66045, USA}

\author[0000-0002-4407-885X]{Alyssa Bulatek}
\affiliation{Department of Astronomy, University of Florida, PO Box 112055, Gainesville, FL 32611 USA}

\author{Michael Wieber}
\affiliation{Department of Physics and Astronomy, University of Kansas, 1251 Wescoe Hall Drive, Lawrence, KS 66045, USA}

\author[0000-0002-8586-6721]{Pablo Garc\'ia}
\affiliation{Chinese Academy of Sciences South America Center for Astronomy, National Astronomical Observatories, CAS, Beijing 100101, China}
\affiliation{Instituto de Astronom\'ia, Universidad Cat\'olica del Norte, Av. Angamos 0610, Antofagasta, Chile}

\author[0000-0001-5389-0535]{Denise Riquelme-V\'asquez}
\affiliation{Departamento de Astronom\'ia, Universidad de La Serena, Ra\'ul Bitr\'an 1305, La Serena, Chile}

\author[0000-0003-1337-9059]{Xing Pan}
\affiliation{School of Astronomy and Space Science, Nanjing University, 163 Xianlin Avenue, Nanjing 210023, P.R.China}
\affiliation{Key Laboratory of Modern Astronomy and Astrophysics (Nanjing University), Ministry of Education, Nanjing 210023, P.R.China}
\affiliation{Center for Astrophysics | Harvard \& Smithsonian, 60 Garden Street, Cambridge, MA 02138, USA}

\author[0000-0002-5094-6393]{Jens Kauffmann}
\affiliation{Haystack Observatory, Massachusetts Institute of Technology, 99 Millstone Road, Westford, MA 01886, USA}

\author[0000-0002-7269-342X]{Marc W. Pound}
\affiliation{University of Maryland, Department of Astronomy, College Park, MD 20742-2421, USA}

\author[0000-0003-4019-0673]{Enrico Di Teodoro}
\affiliation{Department of Physics and Astronomy, Johns Hopkins University, Baltimore, MD 21218, USA}

\author[0000-0003-2133-4862]{Thushara Pillai}
\affiliation{Haystack Observatory, Massachusetts Institute of Technology, 99 Millstone Road, Westford, MA 01886, USA}

\author[0000-0001-6199-9848]{Emad Alkhuja}
\affiliation{Max-Planck-Institut f\"ur Radioastronomie, Auf dem H\"ugel 69, 53121 Bonn, Germany}
\affiliation{Astronomy Department, Faculty of Science, King Abdulaziz University, P.O. Box 80203, Jeddah 21589, Saudi Arabia}

\author[0000-0002-8455-0805]{Yue Hu}
\affiliation{Institute for Advanced Study, 1 Einstein Drive, Princeton, NJ 08540, USA}

\author{Dennis Perlot}
\affiliation{University of Connecticut, Department of Physics, 196A Auditorium Road Unit 3046, Storrs, CT 06269 USA}

\author{Howard A. Smith}
\affiliation{Center for Astrophysics | Harvard \& Smithsonian, 60 Garden Street, Cambridge, MA 02138, USA}

\author[0000-0002-2887-5859]{Victor M. Rivilla}
\affiliation{Centro de Astrobiolog\'{\i}a (CSIC-INTA), Ctra Ajalvir km 4, 28850, Torrej\'{o}n de Ardoz, Madrid, Spain}

\author[0000-0001-9822-7817]{Wenyu Jiao}
\affiliation{Shanghai Astronomical Observatory, Chinese Academy of Sciences, 80 Nandan Road, Shanghai 200030, P.\ R.\ China}

\author[0000-0001-5950-1932]{Fengwei Xu}
\affiliation{Max-Planck-Institut f\"ur Radioastronomie, Auf dem H\"ugel 69, 53121 Bonn, Germany}
\affiliation{Kavli Institute for Astronomy and Astrophysics, Peking University, Beijing 100871, People's Republic of China}

\author[0000-0001-9656-7682]{Jonathan D. Henshaw}
\affiliation{Max-Planck-Institut f\"ur Astronomie, K\"onigstuhl~17, D-69117 Heidelberg, Germany}

\author[0000-0001-6431-9633]{Adam Ginsburg}
\affiliation{Department of Astronomy, University of Florida, PO Box 112055, Gainesville, FL 32611 USA}

\author[0000-0001-8782-1992]{Elisabeth A.C. Mills}
\affiliation{Department of Physics and Astronomy, University of Kansas, 1251 Wescoe Hall Drive, Lawrence, KS 66045, USA}

\author[0000-0003-2384-6589]{Qizhou Zhang}
\affiliation{Center for Astrophysics | Harvard \& Smithsonian, 60 Garden Street, Cambridge, MA 02138, USA}

\author[0000-0001-6353-0170]{Steven N. Longmore}
\affiliation{Astrophysics Research Institute, Liverpool John Moores University, IC2, Liverpool Science Park, 146 Brownlow Hill, Liverpool L3 5RF, UK}
\affiliation{Cosmic Origins Of Life (COOL) Research DAO, \href{https://coolresearch.io}{https://coolresearch.io}}

\author[0000-0001-9657-8728]{Nguyen M. Khang}
\affiliation{Astrophysics Research Institute, Liverpool John Moores University, IC2, Liverpool Science Park, 146 Brownlow Hill, Liverpool L3 5RF, UK}

\author{Galaxy Salo}
\affiliation{Astrophysics Research Institute, Liverpool John Moores University, IC2, Liverpool Science Park, 146 Brownlow Hill, Liverpool L3 5RF, UK}

\author{Jakub Poznanski}
\affiliation{University of Connecticut, Department of Physics, 196A Auditorium Road Unit 3046, Storrs, CT 06269 USA}

\author[0000-0002-5811-0136]{Dylan P\'are}
\affiliation{Joint ALMA Observatory, Alonso de Cordova 3107, Vitacura, Casilla 19001, Santiago de Chile, Chile}
\affiliation{National Radio Astronomy Observatory, 520 Edgemont Road, Charlottesville, VA 22903, USA}

\author[0000-0002-6362-8159]{Maya A. Petkova}
\affiliation{Department of Physics and Astronomy, Chalmers University of Technology, SE-412 96 Gothenburg, Sweden}

\author[0000-0003-4493-8714]{Izaskun Jim\'enez-Serra}
\affiliation{Centro de Astrobiolog\'ia (CAB), CSIC-INTA, Ctra. de Ajalvir Km. 4, 28850, Torrej\'on de Ardoz, Madrid, Spain}

\author[0000-0003-2619-9305]{Xing Lu}
\affiliation{Shanghai Astronomical Observatory, Chinese Academy of Sciences, 80 Nandan Road, Shanghai 200030, P.\ R.\ China}
\affiliation{State Key Laboratory of Radio Astronomy and Technology, A20 Datun Road, Chaoyang District, Beijing, 100101, P.\ R.\ China}

\author[0000-0001-9281-2919]{Sergio Mart\'in}
\affiliation{European Southern Observatory, Alonso de C\'ordova, 3107, Vitacura, Santiago 763-0355, Chile}
\affiliation{Joint ALMA Observatory, Alonso de C\'ordova, 3107, Vitacura, Santiago 763-0355, Chile}

\author[0000-0002-3078-9482]{\'Alvaro S\'anchez-Monge}
\affiliation{Institut de Ci\`encies de l'Espai (ICE), CSIC, Campus UAB, Carrer de Can Magrans s/n, E-08193, Bellaterra, Barcelona, Spain}
\affiliation{Institut d'Estudis Espacials de Catalunya (IEEC), E-08860, Castelldefels, Barcelona, Spain}

\author[0000-0002-6398-7530]{Bijas Najimudeen}
\affiliation{Jodrell Bank Center for Astrophysics, School of Physics and Astronomy, University of Manchester, Oxford road, United Kingdom, M13 9PL.}

\author[0000-0002-0706-2306]{Christoph Federrath}
\affiliation{Research School of Astronomy and Astrophysics, Australian National University, 233 Mt Stromlo Road, Stromlo ACT 2611, Australia}

\author[0000-0002-0862-0701]{Tabassum S. Tanvir}
\affiliation{Department of Physics and Astronomy, Iowa State University, 2323 Osborn Dr, Ames, IA 50011}

\author[0000-0001-9155-3978]{Pei-Ying Hsieh}
\affiliation{National Astronomical Observatory of Japan, 2-21-1 Osawa, Mitaka, Tokyo 181-8588, Japan}

\author[0000-0001-8135-6612]{John Bally} 
\affiliation{Center for Astrophysics and Space Astronomy, Department of Astrophysical and Planetary Sciences,
University of Colorado, Boulder, CO 80389, USA} 

\author{Paul Ho}
\affiliation{Institute of Astronomy and Astrophysics, Academia Sinica, 11F of ASMAB, AS/NTU No. 1, Sec. 4, Roosevelt Road, Taipei 10617, Taiwan}
\affiliation{East Asian Observatory, 660 N. A'ohoku, Hilo, Hawaii, HI 96720, USA}

\author[0000-0001-6113-6241]{Mattia C. Sormani}
\affiliation{Department of Physics, University of Surrey, Guildford GU2 7XH, UK}
\affiliation{Universit{\`a} dell’Insubria, via Valleggio 11, 22100 Como, Italy}

\author[0000-0002-9483-7164]{Robin G. Tress}
\affiliation{Institute of Physics, Laboratory for Galaxy Evolution and Spectral Modelling, EPFL, Observatoire de Sauverny, Chemin Pegasi 51, 1290 Versoix, Switzerland}

\author[0000-0002-3972-1978]{Jaime E. Pineda}
\affiliation{Max-Planck-Institut f\"ur extraterrestrische Physik, Gie\ss enbachstra\ss e 1, 85748 Garching bei M\"unchen, Germany}